\documentclass[11pt,preprint]{aastex}

\newcommand{\dF}{{^{^*}\!\!F}}
\newcommand{\bP}{{\bf P}}
\newcommand{\bF}{{\bf F}}
\newcommand{\bU}{{\bf U}}

\newcommand{\sB}{{\mathcal{B}}}
\newcommand{\eps}{\epsilon}
\newcommand{\detg}{{\sqrt{-g}}}
\newcommand{\del}{{\partial}}
\newcommand{\tu}{{\tilde{u}}}
\newcommand{\tQ}{{\tilde{Q}}}
\newcommand{\udB}{{{u_\mu \sB^\mu}}}
\newcommand{\OneDW}{{$\mathrm{1D}_W$ }}
\newcommand{\OneDWc}{{$\mathrm{1D}_W$}}
\newcommand{\OneDWb}{{$\mathrm{\mathbf{1D}}_\mathbf{W}$ }}
\newcommand{\OneDWi}{{$\mathrm{\mathit{1D}}_\mathit{W}$ }}
\newcommand{\OneDVsq}{{$\mathrm{1D}^\star_{v^2}$ }}
\newcommand{\OneDVsqc}{{$\mathrm{1D}^\star_{v^2}$}}
\newcommand{\OneDVsqb}{{$\mathrm{\mathbf{1D}}^\star_\mathbf{v^2}$ }}
\newcommand{\OneDVsqi}{{$\mathrm{\mathit{1D}}^\star_\mathit{v^2}$ }}
\newcommand{\OneDVsqOrig}{{$\mathrm{1D}_{v^2}$ }}
\newcommand{\OneDVsqOrigc}{{$\mathrm{1D}_{v^2}$}}
\newcommand{\OneDVsqOrigb}{{$\mathrm{\mathbf{1D}}_\mathbf{v^2}$ }}
\newcommand{\OneDVsqOrigi}{{$\mathrm{\mathit{1D}}_\mathit{v^2}$ }}

\newcommand{\beq}[1]{\begin{equation} #1 \end{equation}}

\newcommand{\ibanez}{Ib\'{a}\~{n}ez, J.~$\mathrm{M}^{\underline{a}}$}
\newcommand{\marti}{Mart\'{i}, J.~$\mathrm{M}^{\underline{a}}$}

\begin{document}

\title{Primitive Variable Solvers for Conservative General Relativistic Magnetohydrodynamics}

\author{Scott C. Noble}
\affil{Physics Department, University of Illinois, Urbana, IL 61801,  U.S.A.}
\email{scn@uiuc.edu}

\and

\author{Charles F. Gammie}
\affil{Physics Department, University of Illinois, Urbana, IL 61801,  U.S.A.}
\email{gammie@uiuc.edu}

\and

\author{Jonathan C. McKinney}
\affil{Harvard-Smithsonian Center for Astrophysics, Cambridge, MA 02138, U.S.A.}
\email{jmckinney@cfa.harvard.edu}

\and

\author{Luca Del Zanna}
\affil{Dipartimento di Astronomia e Scienza dello Spazio
Universit\`{a}  degli Studi di Firenze, Firenze, Italy}
\email{ldz@arcetri.astro.it}

\shortauthors{Noble et al.}
\shorttitle{Primitive Variable Solvers for GRMHD}

\begin{abstract}

Conservative numerical schemes for general relativistic
magnetohydrodynamics (GRMHD) require a method for transforming between
``conserved'' variables such as momentum and energy density and
``primitive'' variables such as rest-mass density, internal energy, and
components of the four-velocity.  The forward transformation (primitive
to conserved) has a closed-form solution, but the inverse transformation
(conserved to primitive) requires the solution of a set of five
nonlinear equations.  Here we discuss the mathematical properties of the
inverse transformation and present six numerical methods for performing
the inversion.  The first method solves the full set of five nonlinear
equations directly using a Newton-Raphson scheme and a guess from the
previous timestep.  The other methods reduce the five nonlinear
equations to either  one or two nonlinear equations that are solved
numerically.  Comparisons between the methods are made using a survey
over phase space, a two-dimensional explosion problem, and a general
relativistic MHD accretion disk simulation.  The run-time of the methods
is also examined.  Code implementing the schemes is available for
download on the web.

\end{abstract}

\keywords{hydrodynamics --- methods: numerical --- MHD}

\section{Introduction}

It is commonly thought that many astrophysical systems contain
relativistic  plasmas with a dynamically significant magnetic field.
Examples include accreting black holes in black hole binaries, galactic
nuclei, gamma-ray bursts, the cores of massive stars undergoing core
collapse, isolated neutron stars, and neutron stars in binary systems.

As a result, there is currently considerable interest in numerical methods
for integrating the equations of general relativistic
magnetohydrodynamics (GRMHD).  Within the last few years about six GRMHD
schemes have been deployed \citep{kom05,koide,harm,dvh,fragile,duezetal,anton}.

Some of these authors \citep{kom05,koide,harm,duezetal,anton} have adopted a
conservative scheme.  This means that the integrated equations are of
the form
\begin{equation}
\del_t \bU(\bP) = -\del_i \bF^i(\bP) + \mathbf{S}(\bP)
\label{conservative-eq}
\end{equation}
Here $\bU$ is a vector of ``conserved'' variables, such as particle
number density, or energy or momentum density in the coordinate frame,
the $\bF^i$ are the fluxes, and $\mathbf{S}$ is a vector of source terms
that do not involve derivatives of $\bP$ and therefore do not affect the
characteristic structure of the system.  
$\bU$ is conserved in the sense that, if $\mathbf{S} = 0$,
the rate of change of the integral of $\bU$ over the volume depends only
on fluxes at the boundaries, by the divergence theorem.  The vector
$\bP$ is composed of ``primitive'' variables such as rest-mass density,
internal energy density, velocity components, and magnetic field
components.  The fluxes and conserved quantities depend on $\bP$.
Conservative numerical schemes advance $\bU$, then depending on the
order of the scheme, calculate $\bP(\bU)$ once or twice per timestep.

In nonrelativistic conservative MHD schemes the conserved quantities are
trivially related to the primitive variables; both the forward
transformation $\bP \rightarrow \bU$ and the inverse transformation $\bU
\rightarrow \bP$ have a closed-form solution.  In GRMHD (or even SRMHD)
$\bU(\bP)$ is a complicated, nonlinear relation.  The inverse transformation
has no closed-form solution and must be performed numerically.  This
numerical operation must be robust, accurate, and fast---it is at the
heart of all conservative GRMHD schemes.

In this paper we investigate several schemes for the inversion
$\bP(\bU)$ and test each in an axisymmetric simulation of accretion onto
a rotating black hole.  \S~\ref{sec:definitions} covers definitions and
notational matters.  \S~\ref{sec:inversion-schemes} describes five
distinct formulations of the algebraic equations to be solved
numerically, and \S~\ref{sec:numer-impl} describes how each method is
implemented numerically.   \S~\ref{sec:tests} describes the performance
of these methods in the context of a survey over a range of primitive
variable values and in two typical applications.
\S~\ref{sec:conclusion} summarizes our findings and contains a guide to
our results for those wishing to make their own implementation.  We will
assume throughout that $G = c = 1$.

\section{Definitions}
\label{sec:definitions}
Throughout this paper we follow standard notation \citep{mtw}.  
We work in a coordinate basis with metric components $g_{\mu\nu}$ and 
independent variables $t,x_1,x_2,x_3$.
The normal observer's four-velocity is $n_\mu = (-\alpha, 0, 0, 0)$ in this 
coordinate basis, where $\alpha^2 = -1/g^{tt}$ is the square of the
lapse.  

The fluid is described by its four-velocity  $u^\mu$, rest-mass density
$\rho_\circ$, internal energy per unit (proper) volume $u$, and pressure
$p$.  The electromagnetic field is described by the field tensor
$F^{\mu\nu}$, which is antisymmetric, and its dual
\begin{equation}
\dF^{\mu\nu} = {1\over{2}} \eps^{\mu\nu\kappa\lambda} F_{\kappa\lambda}.
\end{equation}
The field tensor has six degrees of freedom (three
components each for E and B), but three are eliminated by the ideal MHD
condition
\begin{equation}
u_\mu F^{\mu\nu} = 0
\end{equation}
which states that the Lorentz force vanishes in the rest-frame of the
fluid.  It is convenient to describe the field using the magnetic field
four-vector
\begin{equation}
\sB^\mu \equiv - n_\nu \dF^{\mu\nu}.
\end{equation}
Notice that $\sB^\mu n_\mu =0$ and the vector $\sB$ differs from the magnetic 
field variables 
\beq{
B^i \equiv \dF^{it} = {\sB^i\over\alpha}
}
that we have used in
earlier papers in this series \citep{harm,gsm,bz}.  It is also
useful to define the projection tensors
\begin{equation}
h_{\mu\nu} = g_{\mu\nu} + u_\mu u_\nu
\end{equation}
which projects into a space normal to the fluid four-velocity $u^\mu$
and
\begin{equation}
j_{\mu\nu} = g_{\mu\nu} + n_\mu n_\nu
\end{equation}
which projects into the space normal to the normal observer $n^\mu$.

The governing equations for GRMHD are conservation of
stress-energy,
\begin{equation}
{T^{\mu}}_{\nu;\mu} = 0
\end{equation}
conservation of particle number,
\begin{equation}
\left(\rho_\circ u^\mu\right)_{;\mu} = 0,
\end{equation}
and the homogeneous Maxwell equations
\begin{equation}
{\dF^{\mu\nu}}_{;\nu} = 0.
\end{equation}
Often, we will assume a $\Gamma$-law gas, with equation of state 
\begin{equation}
p = \left(\Gamma - 1\right) u.  \label{gamma-law-eos}
\end{equation}
Some (but not all) of the schemes described here do not work for other
equations of state.   We will note where our assumed equation of state is
essential.

The GRMHD stress-energy tensor is
\begin{equation}
T^{\mu\nu} = \left(w + b^2\right) u^\mu u^\nu 
+ \left(p + {b^2\over{2}}\right) g^{\mu\nu} - b^\mu b^\nu.
\end{equation}
Here 
\begin{equation}
w = \rho_\circ + p + u,
\end{equation}
$b^2 = b^\mu b_\mu$, and 
\begin{equation}
b^\mu = {1\over{\gamma}}{h^\mu}_\nu \sB^\nu, 
\end{equation}
where 
\begin{equation}
\gamma \equiv - n_\mu u^\mu \ 
\end{equation}
is the Lorentz factor of the flow as measured in the normal observer frame; in our 
coordinate basis $\gamma = \alpha u^t$. 

There are many possible choices of conserved and primitive variables.
For definiteness, we will make a specific choice that is nearly
identical to that used in the HARM code \citep{harm}.  Once an inversion
scheme is found for this particular set many other choices can be
obtained by simple algebraic transformations.

For primitive variables, we use $\rho_\circ, u, \sB^i$, and $\tu^i \equiv
{j^i}_\mu u^\mu$ ($\tu^t = 0$).  These velocities are the projection of
the plasma four-velocity into the space perpendicular to $n_\mu$.  They
can be rewritten $\tu^i = u^i + \alpha \gamma g^{ti}$, or, in the
language of the ADM (Arnowitt-Deser-Misner) formalism $\tu^i = u^i +
\gamma \beta^i/\alpha$, where $\beta^i = \alpha^2 g^{ti}$ is the
``shift''.  The $\tu^i$s are numerically convenient because they range
from $-\infty$ to $\infty$.

For conserved variables, it is helpful to first write out the basic
equations in full.  The equations of energy-momentum conservation are
\begin{equation}
\del_t \left(\detg {T^t}_\mu\right) + \del_i \left(\detg {T^i}_\mu\right) =
	\detg {T^\kappa}_\lambda \Gamma^\lambda_{\nu\kappa},
\end{equation}
where $g \equiv {\rm Det}(g_{\mu\nu})$.
The associated conserved variables are $\detg {T^t}_\mu$.  It is
convenient to convert these to
\begin{equation}
Q_\mu \equiv - n_\nu {T^\nu}_\mu = \alpha {T^t}_\mu  , 
\end{equation}
which is the energy-momentum density in the normal observer frame.

The Maxwell equations yield three evolutionary equations
\begin{equation}
\del_t \left(\detg B^i\right) = -\del_j \left[\detg \left(b^j u^i - b^i u^j\right)\right]
\end{equation}
and the constraint
\begin{equation}
\del_i \left(\detg B^i\right) = 0,
\end{equation}
which does not concern us here.

The particle number conservation equation is 
\begin{equation}
\del_t \left(\detg \rho_\circ u^t\right) = -\del_j \left(\detg \rho_\circ u^j\right) .
\end{equation}
The conserved variable for this equation is $\detg \rho_\circ u^t$.  It is convenient to
convert this to 
\begin{equation}
D \equiv -\rho_\circ n_\mu u^\mu = \rho_\circ \alpha u^t  = \gamma \rho_\circ \ .
\end{equation}  
This is the density measured in the normal observer
frame.

To sum up, the eight conserved variables are $Q_\mu, D$, and $\sB^i$.
We are now in a position to give explicit expressions for the forward
transformation $\bU(\bP)$.

First, we calculate $p = \left(\Gamma - 1\right) u$, $w = \rho_\circ + u + p$, $\gamma
= \sqrt{1 + g_{ij} \tu^i \tu^j}$, 
$u^\mu = \left(\gamma/\alpha, \tu^i - \alpha \gamma g^{ti}\right)$, and 
$b^\mu = {h^\mu}_\nu \sB^\nu/\gamma$.  
Then
\begin{equation}\label{PRIM1}
D = \gamma \rho_\circ ,
\end{equation}
\begin{equation}\label{PRIM2}
Q_\mu = \gamma \left(w + b^2\right) u_\mu - \left(p + b^2/2\right) n_\mu 
+ \left(n_\nu b^\nu\right) b_\mu ,
\end{equation}
and of course the $\sB^i$s are both primitive and conserved variables.
Along the way, it is computationally efficient to use the identities
\begin{equation}\label{bsqexp}
b^2 = {1\over{\gamma^2}} \left[\sB^2 + \left(\sB^\mu u_\mu\right)^2\right]
\end{equation}
and
\begin{equation}\label{ndbeq}
n_\mu b^\mu = - \udB .
\end{equation}
The next section will show how these equations are used to perform the 
inverse transformation. 

\section{Inversion Schemes}
\label{sec:inversion-schemes}

No closed-form inversion of (\ref{PRIM1},\ref{PRIM2}) from $D, Q_\mu$ to
the primitive variables $\rho_\circ, u, \tu^i$ is known, so the primitive
variables must be found numerically.  In this section, we describe and
discuss several methods for solving these equations. 

A popular and robust means of solving systems of algebraic equations
numerically is the Newton-Raphson (NR) scheme (see Section 9.6 of
\cite{nr}).  Since it is also simple to code, we will employ this method
by default.  

\subsection{\OneDWb Scheme}
\label{sec:1d-scheme}

It seems likely that the solution would be obtained more efficiently if
one could manipulate the five fundamental equations to reduce the
dimensionality of the system, as was done by \cite{ldz} for special
relativistic MHD.  Our procedure for reducing the 5D system is to
evaluate certain scalars from the conserved variables, then expand
expressions for these scalars to simplify the solution.  We will use
$D$, $Q_\mu \sB^\mu$, and $Q^\mu n_\mu$, which is the energy density
measured in the normal observer frame.  We will also use $\tQ^2$, where
$\tQ^\nu = {j^\nu}_\mu Q^\mu$, which is the energy-momentum flux
perpendicular to the normal observer.  Ultimately, we obtain two
equations dependent only on the known conserved variables and two
independent variables $\gamma$ and $w$.  Instead of these two unknowns,
however, we use $W\equiv w \gamma^2$ and $v^2\equiv v_i v^i$, where $v^i
= \tu^i / \gamma$ is the flow velocity relative to the normal observer.
These new variables simplify the equations and lead to numerical schemes
that are more robust.  Whenever $\gamma$ is stated from now on, it is
implied that it is calculated from $v^2$ via the identity  $\gamma^2 =
1/\left(1 - v^2\right)$.

A consistent procedure can be developed as follows.  First, expand the
definition of $Q_\mu$ to find the key result
\begin{equation}\label{BdQeq}
\sB^\mu Q_\mu = \left(\udB\right) W / \gamma .
\end{equation}
Since $\sB^\mu Q_\mu$ can be evaluated in terms of the (known) conserved
variables, this equation can be used to eliminate $\udB$ in favor of the
unknowns $v^2$ and $W$.  

Next, expand $\tQ^2$ using (\ref{BdQeq}) to find
\begin{equation}\label{tQeq}
\tilde{Q}^2 = 
v^2 \left(\sB^2 + W\right)^2
- {\left(Q_\mu \sB^\mu\right)^2 \left(\sB^2 + 2 W\right)
   \over{W^2}}.
\end{equation}
Finally, we solve for $v^2$ to find the simple result
\begin{equation}\label{vsqeq}
v^2 = \frac{ \tQ^2 W^2 + \left(Q_\mu \sB^\mu\right)^2 
\left( \sB^2 + 2 W \right) }{ \left(\sB^2 + W \right)^2 W^2 } \  .
\end{equation}
This is the relativistic counterpart of the nonrelativistic expression
$v^2 = P^2/\rho^2$, where $P$ is the magnitude of the momentum density 
vector $\rho {\bf v}$.  It gives us an explicit expression for $v^2(W)$ 
that results in positive definite evaluations if implemented in the code 
as shown here.   

Next, expand $Q_\mu n^\mu$ using (\ref{BdQeq}) and the variables
$\gamma$ and $W$:
\begin{equation}\label{finaleq}
Q_\mu n^\mu = -\frac{\sB^2}{2} \left( 1 + v^2 \right)
+ {\left(Q_\mu \sB^\mu\right)^2\over{2 W^2}} 
- W
+ p(u,\rho_\circ).
\end{equation}
This remaining equation then yields a solution for $W$.  Specifically,
the \OneDWi \emph{scheme} solves one nonlinear algebraic equation
(\ref{finaleq}), which is now only a function of $W$ since (\ref{vsqeq})
is used to eliminate $v^2$. 

Once $v^2$ and $W$ are found one can recover $w$, $\rho_\circ$ (from $D$),
and $u$.  The next step is to find $\tu^i$ using $\tilde{Q}^i$.
After some manipulation one finds
\begin{equation}
\tilde{Q}^\mu = {1\over{\gamma}} \left(W + \sB^2\right) \tu^\mu 
- {\left(u_\nu \sB^\nu\right) \sB^\mu\over{\gamma}}.
\end{equation}
Since $\udB = (\gamma/W) (Q_\mu B^\mu)$, this can be used to solve
for $\tu^i$:
\begin{equation}
\tu^i = \frac{\gamma}{W+\sB^2} \left[ \tilde{Q}^i  
         +   \frac{ \left(Q_\mu \sB^\mu\right)  \sB^i }{W} \right]
\label{u-i}
\end{equation}

\subsection{2D Scheme}
\label{sec:2d-scheme}

The equation to be solved in the \OneDW method includes a quotient of
polynomials in $W$ since it implicitly uses (\ref{vsqeq}) for $v^2$.
This suggests that numerical pathologies might arise near roots.  By
solving the two, simpler equations (\ref{tQeq},\ref{finaleq})
simultaneously for $W$ and $v^2$, one may eliminate such problems.  We
call this method the \emph{2D scheme} since it involves solving a
two-dimensional algebraic system. 

We find that using $v^2$ instead of $\tu^2$ or $\gamma$ is particularly
advantageous for this method. This is because equations
(\ref{tQeq},\ref{finaleq}) are linear only in $v^2$ (modulo the
$v^2$-dependence of the state equation) and not in $\tu^2$ or $\gamma$.
The linear dependence on $v^2$ increases the rate of convergence for
this quantity and is guaranteed to be well-behaved in the vicinity of a
root. 

\cite{koide1} and \cite{koide} also used a
two-dimensional method.  But instead of $v^2$ and $W$, they use
$\left(\gamma-1\right)$ and $\left(\udB\right)$ as independent
variables.  We have not tried their method since the functions they
minimize are more complicated than ours and assume a $\Gamma$-law state
equation.  Further, it is likely that these two methods would perform
similarly since one can eliminate $W$ for $\udB$ in our method via
equation (\ref{BdQeq}).  As mentioned earlier, we find better
performance using $v^2$ instead of $\gamma$. 

\subsection{5D Scheme}
\label{sec:5d-scheme}

The simplest procedure, and the one we used initially in the HARM code,
is to invert the five equations (\ref{PRIM1}) to (\ref{PRIM2}) using a
multidimensional NR scheme.  This requires evaluating the matrix of
derivatives $\del \bU/\del \bP$.  Further, a Newton iteration of this
system involves more operations than the \OneDW and 2D schemes since it
requires calculating elements of $\del \bU/\del \bP$ and a matrix
inversion.  Also, we find that it requires an initial guess that is
close to the solution.  This is almost always available from the last
timestep (and if the guess is not good it usually means something is
wrong).  Because it involves root-finding in a five-dimensional space we
call this the \emph{5D scheme}.  The conserved variables used in this
method are the same as those used in \cite{harm}.  All other methods
described here use $D$ and $Q_\mu$.

Notice that the \OneDWc, 2D, and 5D schemes do not require a particular
equation of state.  For example, derivatives of $p$ with respect to the
independent variables could be obtained from equation of state tables
using finite difference approximations. Two of the next three methods,
however, assume a $\Gamma$-law equation of state. 

\subsection{\OneDVsqOrigb and \OneDVsqb Schemes}
\label{sec:zbl-scheme}

Another scheme can be derived if we assume that the equation of state is 
\begin{equation}
\label{eos}
p \ = \ \left( \Gamma - 1 \right) u \ = \ {{\Gamma - 1}\over\Gamma} 
\left(w - {D\over{\gamma}}\right).
\end{equation}
If we make this substitution and use the expression for $v^2(W)$,
equation (\ref{finaleq}) reduces to an eighth-order polynomial in $W$,
for which there is no general closed-form solution according to the
theorem of Abel-Ruffini.   It is simplest to solve this single nonlinear
equation numerically.  Because of the complexity introduced by the
equation of state there is likely no general way of isolating the
physical root, and one must simply look for a solution that is close to
the solution of the last timestep.  With other state equations, it may
not be possible to express (\ref{finaleq}) in polynomial form
at all. 

It is worth noting how this situation resolves itself in relativistic
\textit{hydro}dynamics.  Equation (\ref{vsqeq}) becomes
\begin{equation}
\tu^2 = {\tQ^2\over{W^2 - \tQ^2}}
\end{equation}
and equation (\ref{finaleq}) becomes
\begin{equation}
Q_\mu n^\mu = p(W,D/\gamma) - W .
\end{equation}
Obviously, there is no general solution since one has the freedom of
choosing an equation of state.  Our particular equation of state,
however, yields a quartic whose solution was discussed by
\cite{eulderink}.

Instead of solving the eighth-order polynomial directly, the
\OneDVsqOrigi \emph{scheme}, which is a modified version of the special
relativistic method described in \cite{ldz}, solves a cubic equation for
$W(\tu^2)$ and a nonlinear equation for $\tu^2$.  The cubic described in
\cite{ldz}, however, can sometimes have two positive, real solutions for
$W$.  The larger root appears to be the physical one, though a general
proof of this has eluded us.  In order to eliminate this
uncertainty, we use the following cubic equation which we have proved
leads to only one, positive solution\footnote{The signs of the cubic
solutions can be derived from the locations of the cubic's local
extrema.  The existence of only one positive solution stems from the
following properties of the cubic: 1) $\partial C /\partial W|_{W=0} =
0$; 2) $C(W=0)\le0$ always; and 3) $\lim_{W\rightarrow\pm\infty} C(W) =
\pm \infty$.}   
\beq{
C(W) = W^3 + 3 d_2 W^2 - 4 d_0  = 0  \label{new-cubic}
}
where 
\beq{
d_0 \equiv \frac{ \Gamma \left( Q_\mu \sB^\mu \right)^2 }{ 8 \left[ 1 
+ v^2 \left( \Gamma - 1 \right) \right] } \ ,
\label{d0}
}

\beq{
d_2 \equiv \frac{ \Gamma }{ 3 \left[ 1 + v^2 \left( \Gamma - 1 \right) \right] } \left[ 
Q_\mu n^\mu  + \frac{\sB^2}{2} \left( 1 + v^2 \right) 
+ D \left( 1 - 1/\Gamma \right) \sqrt{1-v^2}  \right] \ .
\label{d2}
}

Equation~(\ref{new-cubic}) results from  multiplying equation
(\ref{finaleq}) by $\Gamma W^2 / \left[ 1 + v^2 \left(\Gamma -
1\right)\right]$ and substituting equation (\ref{eos}) for
$p(W,D/\gamma)$.  In general, there are three solutions which are given
in closed-form by Cardano's formula \citep{cubic,beta}.  If our cubic
has a positive and real solution then there is only one positive and
real solution, and it can be shown that this solution is always equal to
the following
\beq{
W = - d_2  + \left( 2 d_0 - d_2^3 + \sqrt{\mathcal{D}} \right)^{1/3} 
           + \left( 2 d_0 - d_2^3 - \sqrt{\mathcal{D}} \right)^{1/3} \ , 
\label{gen-cubic-sol}
}
where 
\beq{
\mathcal{D} \, \equiv \, 4 d_0 \left( d_0 - d_2^3 \right)  \ . \label{D-cubic}
}
It is useful to know that $d_0 \ge 0$ always and $d_2$ can take negative and positive values. 
When $\mathcal{D}<0$, the solution can be expressed in a simpler form:
\beq{
W(\mathcal{D}<0) \ = \ d_2 \left[ \cos\left(\theta/3\right) - 1 \right] \quad , 
\quad \theta = \cos^{-1} \left[2d_0/d_2^3 - 1\right]
\ .  \label{simp-cubic-sol}
}
Table~\ref{table:phys-solutions} lists the possible physical solutions of (\ref{new-cubic}) 
depending on the particular values of $d_0$ and $d_2$. 
In the \OneDVsqOrig scheme, we use the physical solution for $W(v^2)$
and numerically solve an equation proportional to equation (\ref{tQeq})
for $v^2$.   That is, \OneDVsqOrig solves for $v^2$ via a NR method in
which the physical cubic solution is calculated for each iteration. 

One drawback of this method is that it requires that the state equation
be linear in $W$.  Solving equation (\ref{finaleq}) numerically instead
makes the method compatible with general equations of state.  We call
this technique the \OneDVsqi \textit{scheme}.  It consists of taking NR
iterations to find $W(v^2)$ nested within Newton iterations to find
$v^2$.  In other words, for each Newton update of $v^2$, we solve
equation (\ref{finaleq}) for $W$ using a separate NR method assuming the
most current value for $v^2$.  This supplies the next $v^2$ iteration
with a consistent value for $W$.  Surprisingly, this  nested NR method
(\OneDVsqc) is faster, more robust and more accurate than the
\OneDVsqOrig method.  The difference in accuracy and efficiency is
likely due to the appearance of transcendental functions and condition
statements in the closed-form solution of (\ref{new-cubic}).

\subsection{Polynomial Scheme}
\label{sec:poly-scheme}

The substitution of equation (\ref{vsqeq}) into equation (\ref{finaleq})
leads to an eighth-order polynomial in $W$ if we assume a $\Gamma$-law
equation of state (\ref{eos}).  This suggests the possibility of using a
general polynomial root-finding method (such as Numerical Recipes' {\tt
zroots}) that finds all 8 roots.  We will call this the {\it polynomial
scheme}.

The physical root can be identified by requiring that it also solve the
five equations $\bU = \bU(\bP)$.  Unfortunately this test can sometimes
yield ambiguous results due to amplification of roundoff error, making
it difficult to identify the correct solution.  This method also turns
out to be computationally expensive.

\section{Numerical Implementations}
\label{sec:numer-impl}

In the previous section, we described the mathematical framework that
embodies the primitive variable inversion methods we have tested.  We
will now discuss the details of the numerical methods we have used to
test the performance of these formulations.  All routines (in the C
language) are available for download from the web.\footnote{The current 
version of the source code for each method described in this paper can be 
found in the electronic edition of the Journal, 
while a maintained version can be obtained at
\url{http://rainman.astro.uiuc.edu/codelib/codes/pvs\_grmhd/}.
The methods can be used with any user-supplied spacetime metric.}

We use the Newton-Raphson scheme to solve the nonlinear algebraic
equations of the \OneDWc, 2D, 5D, \OneDVsqc, and \OneDVsqOrig methods.
An excellent description of this method---as well as other root-finding
methods---can be found in \cite{nr}; we will only state unique aspects
of our implementation here and defer further explanation to our source
code.  Let $\mathbf{R}$ denote the system of nonlinear equations, or
``residuals,'' $\mathbf{x}$ the independent variables for which we are
solving, and $\mathbf{J}$ the Jacobian whose $(i,j)$-component is
$\partial \mathrm{R}_i / \partial x_j$.  The NR procedure assumes that
$\mathbf{R}$ is smooth and nearly linear, so that we can make
consecutive linear corrections to our guess that lead us toward the
root.  This \emph{NR step} is defined as the following:
\beq{
\Delta \mathbf{x} = - \mathbf{J}^{-1} \cdot \mathbf{R}  \quad . \label{nr-step}
}

We measure convergence using a ``Newton-Raphson error function,''
$E_\mathrm{NR} = \left|\Delta W / W\right|$, where $\Delta W$ is the
change in $W$ between the two most recent NR iterations.  We find that
other convergence criteria yield little if any improvement, which is not
surprising since $W$ is the crucial variable.  In particular, a
convergence criterion based on residual errors in the solution of
equations (\ref{PRIM1},\ref{PRIM2}) is not used for any of the schemes
since it is difficult to normalize the error
properly \textit{a priori} over the entire parameter space.  

The condition for stopping the iterative procedure involves three
parameters: $\mathtt{TOL}$, $\bar{N}_\mathrm{NR}$ and
$N_\mathrm{extra}$.  If $E_\mathrm{NR}$ falls below $\mathtt{TOL}$ after
performing at least one iteration, then only $N_\mathrm{extra}$ more
iterations are performed.  A solution is said to be found if the
tolerance criterion is satisfied before $\bar{N}_\mathrm{NR}$
iterations.  

The impetus for the additional $N_\mathrm{extra}$ iterations is one of
efficiency and accuracy.  When the error reaches our tolerance level,
the extra iterations try to reduce it even further.  Since this can
sometimes be fruitless---e.g. the solution error becomes insensitive to
subsequent NR steps---we set $N_\mathrm{extra}$ to a small number.  The
parameters used by default in this paper are $\mathtt{TOL}=10^{-10}$ and
$N_\mathrm{extra}=2$, which should yield solutions accurate to within
roundoff error if the guess is sufficiently near the true solution,
since the NR method converges quadratically.

What value should be chosen for $\mathtt{TOL}$?  Ideally the inversion
error $\mathtt{TOL}$ should be smaller than the truncation error that is
inevitably introduced in the numerical evolution of $\bU$.  But
truncation error is difficult to estimate on-the-fly \footnote{Local
truncation error estimation \emph{can} be accomplished in place during a
simulation without knowing the exact solution beforehand by solving the
equations of motion additionally on a ``shadow'' grid whose grid spacing
is half that of the base grid.  The truncation error estimate is then
calculated  by comparing the two solutions at coincident timesteps
\citep{bo,matt}.} and cannot usually be estimated \textit{a priori}.  

Different problems may require different values of $\mathtt{TOL}$.  One
class of problems that is particularly sensitive to the value of
$\mathtt{TOL}$ is the evolution of small amplitude waves.  This is a
problem of some interest, as it is frequently used in convergence
testing numerical methods.  In this case the fluctuating part of the
fluid variables is small, and the truncation error is a small fraction
of that.  The truncation error can therefore be a very small fraction of
the conserved variables, and $\mathtt{TOL}$ must be comparably small to
avoid spoiling convergence.  On the other hand, the measurements of the
accretion rates of rest mass, energy, and angular momentum in the
relativistic disk simulation discussed below appear to be peculiarly
{\it insensitive} to $\mathtt{TOL}$.  The only safe course is to
directly check the sensitivity of numerical measurements to
$\mathtt{TOL}$.

During a sequence of NR iterations, an independent 
variable may leave its allowed domain (e.g. superluminal velocities).  
In order to prevent numerical divergences and 
unwanted imaginary parts, we reset the independent variable to a value 
in its physical domain.  For example, in the \OneDWc, \OneDVsqOrigc, \OneDVsqc,
2D and polynomial methods, we demand that $0<v^2<1$ and $W>0$.  
No constraints are used in the 5D method, since
experimentation with this has led us to believe that constraining 
$\rho_\circ,u>0$ only increases the likelihood of not converging to a solution.

A special note on the polynomial scheme, where we use Laguerre's method
to find $W$: there is an additional numerical criterion that must be
used for evaluating which of the 8 roots is the physical root.  One
first eliminates roots that have imaginary components above some
threshold value, then evaluates the residuals in the solution of the
basic equations for the remaining roots.  The root with the smallest
residuals is identified as the physical root.  As for the NR methods,
parameters $\mathtt{TOL}$, $\bar{N}_\mathrm{NR}$ and $N_\mathrm{extra}$
are used to control the accuracy of the solutions.

A few final comments on numerical implementation.  In terms of
complexity, the \OneDWc, \OneDVsqc, 2D, and polynomial methods are the
easiest to implement; some care is required in finding the physical
solution to the cubic equation in the \OneDVsqOrig scheme.  The
lower-dimensional NR schemes are simple enough to derive the closed-form
expressions for $\mathbf{R}$, $\mathbf{J}$ and $\Delta \mathbf{x}$ by
hand, so we have coded those in directly.  The 5D scheme, by contrast,
is complicated by the need to evaluate the $25$ elements of
$\mathbf{J}$.  This can be done analytically (this procedure is
time-consuming and prone to error) or numerically using finite
differences (this is simple but introduces additional numerical noise).

\section{Tests}
\label{sec:tests}

The space of possible numerical approaches to the inversion problem is
large, so we can make no claim that any of our methods are optimal or
near-optimal.  In this section, through a series of tests, we show that
some of the methods are ``good enough'' in the sense that (1) they can
be shown not to contribute significantly to the numerical error in
actual astrophysical applications, and (2) some of them are efficient
enough that they do not contribute significantly to the computational
cost of an evolution.  We have expended significant effort in optimizing
each scheme's speed and accuracy, so that these tests make a fair
comparison between the various methods.

We consider three tests.  The first is a parameter space survey in which
the primitive variables are varied over many orders of magnitude and the
accuracy of the solution is evaluated.  The second test places the
inversion routine in a special relativistic MHD code and evolves a
magnetized, cylindrical explosion problem due to Komissarov.  The third
test places the inversion routine in a general relativistic MHD code and
evolves a magnetized disk around a rotating black hole.  Throughout we
assume a the $\Gamma$-law equation of state (\ref{eos}) with
$\Gamma=4/3$.

\subsection{Parameter Space Survey}
\label{sec:param-space-surv}

The inversion routine takes as arguments the 8 conserved variables
(three are magnetic field components and are trivially converted to
primitive variables), guesses for the 5 nontrivial primitive variables,
and the 10 components of the metric.  This parameter space is too large
to cover exhaustively, so we only vary $\rho_\circ$, $u$,
$B^2$, $\gamma$ (the Lorentz factor), and $\Phi \equiv \cos^{-1}\left(
\tu_i B^i / \sqrt{\tu_i \tu^i B_j B^j} \right)$ over the region
spanned by a typical accretion disk evolution:
\beq{
\begin{array}{c}
\log_{10} \rho_\circ \in \left[ -7, 1 \right]  \ , \ \log_{10} u \in \left[ -10, 0 \right] \ , \ 
\log_{10} \gamma \in \left[ 0.002, 2.9 \right] \ , \\[0.25cm]
\log_{10} B^2 \in \left[ -8, 1 \right] \ , \ \cos \Phi \in \left[ -1, 1 \right] .
\end{array}
\label{survey-range}
}
We sample uniformly in the logarithm for each variable.  Specifically,
$40 \times 40 \times 20 \times 20 \times 9$ points are evenly sampled
along dimensions $\log_{10} \rho_\circ \otimes \log_{10} u \otimes \log_{10} \gamma
\otimes \log_{10} B^2 \otimes \cos \Phi$, respectively.  In order to 
choose reasonable relative magnitudes between the components of $\tilde{u}^i$ 
and $B^i$ at a given location with respect to the 
Kerr-Schild metric, we select $9$ points from an accretion disk simulation 
(see Section~\ref{sec:accr-disk-evol}) at which $\cos(\Phi) \simeq  -1, \ldots, 1$. 
The specific values of $\cos(\Phi)$, $\tilde{u}^i$, $B^i$ and coordinates used
for the parameter space survey are given in Table~\ref{table:pspace-points}. 
The overall magnitudes of the $\tu^i$ and $B^j$ are set by 
the values of $\gamma$ and $B^2$ at the given parameter space point.

At each point of the parameter space we perform the forward transformation; this
gives a (nearly) exact solution to the inversion.  We then feed the
inverter a guess obtained by multiplying each primitive variable by $1 +
d$, where $d$ is a random value between $-1$ and $1$.  The same sequence
of random values is used for each method.  This turns out to be 
quite a stringent test, particularly when the random value is near $-1$.  
Even though the maximum threshold for the offset may not be large enough 
to model the behavior at strong shocks, it is approximately the maximum seen for 
$\rho_\circ$ and $u$ in the bulk flow of our accretion disk simulations
(see Section~\ref{sec:accr-disk-evol}).  

Our parameter space surveys were made using $\mathtt{TOL}=10^{-10}$,
$\bar{N}_\mathrm{NR}=30$ and $N_\mathrm{extra}=2$; these values were
determined after the fact to be nearly optimal.

Figure 1 shows the normalized error in the internal energy, $\Delta
u/u$, for all the methods.  These accuracy measurements indicate how
strongly roundoff errors are amplified by the scheme.  The error is
averaged over all the parameters except $u$.  Only the points for which
{\it all} the methods converge are included in the 
average\footnote{Since the 5D method fails at almost half the original points 
and a majority of these failures occur when $\gamma$ is large, this may 
introduce a bias.  We have performed an identical survey in which 
no points were neglected.  Even though solution rates decrease slightly, 
the rankings in Table~\ref{table:pspace-efficiency} stay the same.  The 
high-$E_\mathrm{NR}$ tails (Figure~\ref{fig:errx-pspace}) extend 
to larger values of $E_\mathrm{NR}$;  the $E_\mathrm{NR}$ distributions 
of the 2D, \OneDWc, \OneDVsqc, \OneDVsqOrig and 5D methods now extend to 
$\sim 10^{-10}$, $10^{-9}$, $10^{-9}$, $>1$ and $>1$, respectively.
More iterations are required and all methods but the 2D method have  
$N_\mathrm{NR}$ distributions extending to $N_\mathrm{NR}=30$.  The 2D 
method remains the best method. }.  
Two methods
stand out as less accurate: the polynomial scheme, and the \OneDVsqOrig
scheme.  An examination of $\Delta u/u$ over the entire parameter space
indicates that the the \OneDWc, 2D and \OneDVsq methods yield almost
indistinguishably accurate solutions.  If one ignores failures points,
the 5D method is often as accurate as these methods.  The \OneDVsqOrig
is the fifth most accurate (ignoring failure points), and the polynomial
scheme is the least accurate.  

The relative errors in $\rho_\circ$ and $\tilde{u}^i$ increase 
with increasing $B^2/\rho_\circ$, $B^2/u$,  and $\gamma$ for the most 
accurate schemes.   The relative error of $u$---for the same methods---grows 
with $\rho_\circ/u$, $\gamma$ and $B^2/u$, and
is fairly independent of $B^2/\rho_\circ$.  We find that there is
typically a maximum value of $\rho_\circ / u$ above which the methods nearly 
always fail; this maximum value starts large ($\sim 10^{17}$) 
but reaches unity by the time $\gamma \sim 10^6$ 
\footnote{These numbers were 
obtained in an extended parameter space survey in which we sampled $40^4$ 
points for each of the $9$ values of $\Phi$ over the space
$\log_{10}(\rho_\circ) , \log_{10}(u) , \log_{10}(B^2) \in \left[ -10, 10 \right],$ 
and $\log_{10}(\gamma) \in \left[ 0, 6 \right]$.}.  
We find that poor accuracy ($\gtrsim 1\%$ relative error) in $\rho_\circ$ and 
$\tilde{u}^i$ usually occurs for points with $B^2/\rho_\circ , B^2/u \gtrsim 10^{10}$, 
but $u$ is accurate for $B^2/u \lesssim 10^{12}$. 
The dependency on $\gamma$ seems to not be strongly tied to any of the other 
variables, and we find that---on average---$\gtrsim 1\%$ relative errors are 
seen when $\gamma \gtrsim 10^{3} - 10^{6}$.  Also, the
accuracy of all the methods improves as $\cos \Phi \rightarrow 0$, on
average.  Note that the precise values of these thresholds are not universal 
to all situations, and are quantitatively dependent on all the 
degrees of freedom, including the offset $d$. 

The inversion routine is said to ``fail'' if it has not converged after
$\bar{N}_\mathrm{NR}=30$ iterations.  Failure rates for the methods are
given in Table 2.  Here the 2D method stands out as the best, failing
only $5$ times in the survey of $5.76 \times 10^6$ points.  The \OneDW and
\OneDVsq methods also have low failures rates.  The \OneDVsqOrig and
polynomial schemes fail much more frequently, at a rate
about an order of magnitude lower than that of the least robust 5D method.
Notice that while the polynomial method converges more
often than the 5D method, it often converges to an inaccurate result.

We have found that all the methods asymptote to a minimum failure 
rate as the range in $d$ is diminished, i.e. as $\delta_1 \rightarrow 0$ 
when $d \in [-\delta_1, \delta_1]$.  The 5D method's asymptotic regime ends at 
approximately $\delta_1 \simeq 10^{-3}$, while the other methods are 
insensitive to changes for $\delta_1 < 1$, where $\delta_1=1$ is the maximum
value for which the $\rho_\circ$ and $u$ guesses remain non-negative. 
If we instead vary the overall magnitude of the guess, i.e. 
$\left( 1 + d \right) \rightarrow \delta_2 \left( 1 + d \right)$, set $\delta_1=1$,
and set $\delta_2 \in [ 10^{-6} , 10^6 ]$, 
we find that the 5D method has a minimum failure rate at about $\delta_2 \sim 1$ as 
expected.  The other methods fail at different, yet constant, 
minimum rates up until $\delta_2 \gtrsim 100$;  this may be because the residuals 
used in these routines steepen as $W \rightarrow \infty$.  
This does not imply that every time the density jumps by $\sim 10^3$ the method will 
fail, since the relative offsets are not likely to be large for \emph{all} primitive 
variables at the same time in practice. 
The high failure rate and greater sensitivity to the guess' offset seen with the 5D 
method is easily attributed to the fact that higher-dimensional minimization
problems are more difficult than lower-dimensional problems (e.g. \cite{nr}); 
more dimensions admit residuals with more complicated landscapes and make a method 
more likely to be mired by local minima. 

As another means of demonstrating the accuracy of each method's
solutions, we show histograms in Figure~\ref{fig:errx-pspace} of the
parameter space points binned against the method's exit value of
$E_\mathrm{NR}$.  Any points for which $E_\mathrm{NR}\le10^{-10}$ are
considered acceptable solutions; since only those points for which
solutions are found by all methods are included, there are no points
beyond  $E_\mathrm{NR}=10^{-10}$.  Those methods that explicitly solve
for $W$ (i.e. 2D and \OneDWc) give rise to a power-law distribution
extending up to machine precision since the change in $W$ between
iterations $\Delta W$ is explicitly calculated during the NR step.  The
other methods result in non-zero $\Delta W/W$ only when it is above
machine precision since they calculate it by subtracting the previous
iteration's value from the present one.  From a numerical perspective,
anything below machine precision should be considered equivalent to zero
anyway, so we shall only concern ourselves with the distributions above
machine precision.  In this regime, the 2D method again yields the best
results, followed by \OneDVsqc, \OneDWc, \OneDVsqOrig and 5D in order of
increasing average $E_\mathrm{NR}$.  

What is the maximum accuracy attainable by each method?  We have
surveyed parameter space using values of $\mathtt{TOL}=10^{-15}$ and
$\bar{N}_\mathrm{NR}=200$, which should give close to maximum accuracy.
The distributions of $E_\mathrm{NR}$ for the \OneDWc, 2D and \OneDVsq
methods remained nearly unchanged from those shown in
Figure~\ref{fig:errx-pspace} which used the default (less accurate) set
of parameters; the additional iterations failed to significantly reduce
solution error, on average, for these methods.  The 5D method, however,
does benefit from a larger $\bar{N}_\mathrm{NR}$.  Its accuracy and
failure rate improve the longer it is allowed to iterate, although it
still underperforms the 2D, \OneDW and \OneDVsq solvers.  

In order to measure how quickly the methods converge to a solution, we
have plotted histograms in Figure~\ref{fig:niter-pspace} of the number
of NR iterations, $N_\mathrm{NR}$, each method performed before exiting
per parameter space point.   From the figure, we see that the \OneDWc, 
2D and \OneDVsq methods all typically use between $6-10$ iterations, 
and that they give rise to nearly Gaussian distributions in this
regime. The \OneDVsqOrig method has an extended power-law tail, while
the 5D method has a nearly constant distribution over $N_\mathrm{NR}$.
Even though the \OneDVsqOrig  scheme has a peak at small $N_\mathrm{NR}$
that is narrower and taller than the other schemes---suggesting that it
may be more efficient--- it fails to find a solution quite frequently
and has a substantial tail.  The \OneDW method appears to have the best
convergence rate over the sampled parameter space since it has a steep
exponential tail and it usually converges to a solution in less than
$10$ iterations.

Which scheme is fastest?  Table~\ref{table:pspace-efficiency} shows the
number of solutions per second achieved by each scheme in the parameter
survey, on our 3.06 GHz Intel box.  The two fastest methods---\OneDW and
2D---are about $30\%$ faster than \OneDVsqOrig method, which also has
the smallest average $N_\mathrm{NR}$.  Since the \OneDVsq method
performs many more computations (because of its nested NR scheme), it
takes fourth place.  The 5D method is almost an order of magnitude
slower, due to its larger linear system and larger average
$N_\mathrm{NR}$.  The polynomial method is a factor of 2 slower
than the 5D.

We have tried a variety of schemes for improving the accuracy and
convergence rate of each scheme.  For the NR schemes, we tried
implementing a line searching method, which attempts to find an optimal
step size along the direction of steepest descent \citep{nr}.  This
method saves a few of the nonconvergent outliers in the parameter space
survey for the \OneDW method but does not help anywhere else for any
other method.  We tried this improved \OneDW scheme in an accretion disk
simulation and found that it did not improve the success rate at all in
the disk simulation.  

To sum up, the 2D method has the lowest failure rate and is only
slightly more computationally expensive than the fastest method.  If the
initial guesses are close enough to the solution, the 5D and
\OneDVsqOrig methods fare nearly as well as the more successful \OneDW,
2D, and \OneDVsq methods in accuracy but because of their larger rates
of nonconvergence neither is recommended.  The polynomial method is slow
and inaccurate, and we do not recommend it.  We will ignore the worst of
these methods in the next two sections and discuss the performance of
only four methods: 2D, \OneDWc, \OneDVsqc, and 5D. 


\subsection{Cylindrical Explosion}
\label{sec:cyl-exp}

The parameter space survey may emphasize different aspects of the
inversion routine than a realistic RMHD problem.  Here we consider
inversion routine performance in a special relativistic, slab-symmetric,
magnetized explosion.  We use the same initial conditions as
\cite{kom99}.  The computational domain is defined in Cartesian
coordinates $(x,y) \in [-6,6]\times[-6,6]$ discretized by $200\times200$
points.  A pressure-enhanced cylinder is centered on the origin with
radius $r=0.8$, and is matched exponentially to a constant background
that starts at $r=1$.  The inner state has $\rho_\circ(r<0.8)=10^{-2}$
and $P(r<0.8)=1$, while the outer state is initialized with
$\rho_\circ(r>1)=10^{-4}$ and $P(r>1)=3\times10^{-5}$.  All spatial
components of the velocity are zero at $t=0$, and the magnetic field is
uniform, pointing in the $x$-direction: $\sB^\mu =
\left(0,0.1,0,0\right)$.  

We evolved the initial state using a version of the HARM \citep{harm}
code.  We use an HLL \citep{hll} flux, and linear (second order)
reconstruction with a minmod slope limiter.  We use a Courant factor of
$0.1$ in the evolution.  Figure~\ref{fig:cyl-exp} is a reconstruction of
Figure~10 of \cite{kom99} using our own evolution; the results are
qualitatively identical. 

All the inversion schemes perform similarly on this problem.  Table 3
outlines the speed and failure rates obtained with each method.    When
a solution is found, it is found by all methods in less than $10$
iterations.  The only significant differences are in the run time: the
2D is clearly fastest, while 5D is a factor of 2 slower than the next
fastest method.  The failure rates are the same, because the methods
fail at the same points where the $\bU$ calculated from the evolution is
unphysical.  The small Courant factor may play a role in the low failure
rate here, since smaller timesteps imply smaller changes in the $\bU$
over a timestep, so the initial guess used in the inversion routine
($\bP$ from the previous timestep) is closer to the solution.

\subsection{Accretion Disk Evolution}
\label{sec:accr-disk-evol}

A more challenging context for testing the inversion routines is the
evolution of a magnetized, centrifugally supported torus around a Kerr
black hole (see, e.g., \cite{bz} and \cite{gsm} for similar
applications).  In this particular incarnation of the problem we evolve
a \cite{fm} torus containing a weak poloidal seed field around a Kerr
black hole with $a/M = 0.9375$.  The initial state is unstable to the
magnetorotational instability \citep{mri}, so turbulence develops in the
disk and material accretes onto the black hole.

The inner edge of the initial torus lies as $6 G M/c^2$ and the pressure
maximum lies at $12 G M/c^2$.  The Fishbone-Moncrief equilibrium 
condition of $u_\phi u^t = 4.28$ is used.  On top of the Fishbone-Moncrief 
torus we add a weak magnetic field with vector potential $A_\phi =
\mathrm{Max}\left(\rho_\circ / \rho_\mathrm{max} - 0.2, 0\right)$ where
$\rho_\mathrm{max}$ is the maximum of the disk's rest-mass density.  The
magnetic field amplitude is normalized so that ratio of gas to magnetic
pressure within the disk has a minimum of $100$.  With the addition of
the field the disk is no longer strictly in equilibrium, but it is only
weakly perturbed.

The initial conditions are evolved using the HARM code.  The flux is a
local Lax-Friedrichs flux, and linear (second order) interpolation is
used with a Monotonized Central limiter.  Since HARM is incapable of
evolving a perfect vacuum, we surround the disk in an artificial
atmosphere, or ``floor'' state, with $\rho_{\circ,\mathrm{atm}} =
10^{-4} (r/M)^{-3/2}$ and $u_\mathrm{atm} = 10^{-6} (r/M)^{-5/2}$.
Also, $\rho_\circ$ and $u$ are set to their floor values if and when
they fall below the floor at any point in the evolution.  

The equations of motion are solved using finite difference
approximations on a discrete domain of so-called Modified Kerr-Schild (MKS)
coordinates.  Instead of the standard Kerr-Schild coordinates
$(t,r,\theta,\phi)$, a uniform discretization of MKS coordinates 
$x^\mu = (t,x^{(1)},x^{(2)},\phi)$ is used where 
\beq{
r = e^{x^{(1)}} \quad ,  \quad
\theta = \pi x^{(2)} + \frac{1}{2} \left( 1 - h \right) \, \sin(2 \pi x^{(2)})
\ . 
\label{mks}
}
This coordinate transformation is intended to concentrate cells near
the equator and at small radius. The cells are centered at coordinates
\beq{ 
x^{(1)}_i = x^{(1)}_\mathrm{in} + \left( i + \frac{1}{2} \right) \Delta x^{(1)}
\quad , \quad 
x^{(2)}_j = \left( j + \frac{1}{2} \right) \Delta x^{(2)}
\label{x12i}
}
where $i \in \left[0,N_1-1\right]$, $j \in \left[0,N_2-1\right]$, and
\beq{
x^{(1)}_\mathrm{in} = \log\left(r_\mathrm{in}\right) 
\quad , \quad
\Delta x^{(1)} = \frac{1}{N_1} \log\left( \frac{r_\mathrm{out}}{r_\mathrm{in}} \right)
\quad , \quad 
\Delta x^{(2)} = \frac{1}{N_2}  \quad .
\label{dx12}
}
$N_1$ and $N_2$ are the number of cells along the $x^{(1)}$ and
$x^{(2)}$ axes, respectively; for the $128^2$ ($256^2$) run,  $N_1=N_2=128$ 
($N_1=N_2=256$).   $r_\mathrm{in}$ is chosen so that
approximately ten cells lie behind the event horizon:  
$r_\mathrm{in} \simeq 1.0035 M$ for the $128^2$ run, and  $r_\mathrm{in}
\simeq 1.1702 M$ for the $256^2$ run. 
The remaining parameters are always set to $r_\mathrm{out}=40M$ and
$h=0.3$. 

We have run this torus problem using the 2D, \OneDWc, \OneDVsqc, and 5D
inversion schemes.  The parameters are the same as before:
$\{\mathtt{TOL}, \bar{N}_\mathrm{NR}, N_\mathrm{extra}\}=\{ 10^{-10},
30, 2\}$.  The resolution is $256^2$.  The initial conditions in each
run also contain identical low level noise to initiate the growth of the
magnetorotational instability.  The accretion rates of rest-mass,
energy, and angular momentum for these runs are shown in
Figure~\ref{fig:accretion-rates}.  One would expect that after the disk
becomes turbulent (at $\sim 500~GM/c^3$ in this model) small
differences in the solution due to the inversion routines would be
amplified and that the accretion rates would not track each other very
well.  Evidently the accretion rates are nearly identical until
$900~GM/c^3$, after which their rates follow only vaguely similar
trends.  This suggests that the differences in evolution due to the
inversion routines were small indeed.

The average normalized accretion rates for the internal energy
($\dot{E}/\dot{M}_\circ$), angular momentum ($\dot{L}/\dot{M}_\circ$) and rest-mass
($\dot{M}_\circ$) given in Table~\ref{table:acc-accretion-rates} differ
by only a few percent.  Far larger differences were obtained by changing
the seed used to generate the noise in the initial conditions.  In
addition, the qualitative structure of the disks in steady-state remains
the same at the end of the evolutions.  Shown in
Figure~\ref{fig:log-rho} are snapshots of $\log (\rho_\circ)$ at
$t=2000~GM/c^3$ from the runs using the 2D, \OneDWc, \OneDVsq and 5D
methods (shown left to right, respectively).  The most pronounced
differences are in the corona and funnel regions overlying the bulk of
the disk.  Evidently the solution is stable to changes in the inversion
algorithm.

We have also compared low resolution ($128^2$) runs using the four
methods.  This is interesting because $\bP$ exhibits larger fractional
changes per timestep at lower resolution, so the inversion routines are
put under greater stress.  In Figure~\ref{fig:errx-disk}, we show the
distribution of $E_\mathrm{NR}$ exit values over the course of these
simulations.  The most striking feature of the plot is in the number of
points at which the 5D method fails to converge compared to other
methods.  The 2D method  rarely fails to converge, and the \OneDW and
\OneDVsq methods fail to converge at nearly equal, intermediate rates.
In Table~\ref{table:acc-efficiency} we state the failure rate---i.e. the
frequency at which either an unphysical $\bP$ value is found or no
solution is found---for each $256^2$ run.  The table demonstrates that
the failure rates are approximately the same, suggesting that when the
2D method ``fails'' it is converging to an unphysical $\bP$ solution
instead of failing to converge altogether like the 5D method.  

The inversion schemes are also faster in the disk evolution than in the
parameter space survey.  This is seen in the distribution of
$N_\mathrm{NR}$ over the lower-resolution run shown in
Figure~\ref{fig:niter-disk}.  Many more inversions in the disk run
converge sooner than in the parameter space survey.  The greater
efficiency and accuracy seen in the disk evolution is likely due to the
fact that in the disk evolution the inversion routine is usually
supplied with better initial guesses, from the previous timestep, than
in the parameter space survey.

What is the optimal value of the Newton-Raphson parameters,
$\{\mathtt{TOL}, \bar{N}_\mathrm{NR}, N_\mathrm{extra}\}$, for the most
successful 2D and \OneDW methods?  Almost no new acceptable solutions
are found if we increase $\bar{N}_\mathrm{NR} > 30$, so this is the
natural value for this parameter.  We also reran the disk evolution
using $\mathtt{TOL} =\{10^{-3}$, $10^{-5}$, $10^{-7}$, $10^{-9}$,
$10^{-10}$, $10^{-11}\}$ at a resolution of $256^2$.  Each run was
otherwise identical and each used $\{\bar{N}_\mathrm{NR},
N_\mathrm{extra}\}=\{30,2\}$.  Surprisingly, we discovered that the
evolutions deviated little from each other until \emph{after} the inner
parts of the disks become turbulent, at $t\simeq500 M$.  The relative
difference of any given primitive function between any run and the
$\mathtt{TOL}=10^{-10}$ run grows as $\sim t^{7}$ until it plateaus to a
constant value at $t\gtrsim 1000 M$.  Further, the accretion
rates---i.e. those functions displayed in
Figure~\ref{fig:accretion-rates}---were qualitatively indistinguishable
until $t\simeq800M$.  The total number of NR iterations executed during
the course of a simulation also varied little between these runs. 

Similar runs were made for $\mathtt{TOL} =\{10^{-3}$, $10^{-9}$,
$10^{-10}$, $10^{-11}\}$, but now we reduced the resolution to $128^2$.
Since the timestep is larger in the lower resolution runs, guesses for
the primitive variables are---on average---further away from their
solutions than in the $256^2$ resolution runs.  As a result, the
accretion rates and distribution of $E_\mathrm{NR}$ differed
dramatically between the $\mathtt{TOL}=10^{-3}$ run and the others.
Runs with $10^{-11} < \mathtt{TOL} < 10^{-9}$  have similar accretion
rates, and distributions of $E_\mathrm{NR}$ and $N_\mathrm{NR}$.  This
implies that the guesses in the higher resolution run are so close to
their solutions that performing $3$ iterations---the minimum number of
iterations allowed in the NR schemes when $N_\mathrm{extra}=2$---often
leads directly to an accurate solution independent of \texttt{TOL}.

These tests suggest that the tolerance, if below at least $10^{-3}$, is
inconsequential to our accretion disk simulations when the grid
resolution is sufficiently high and when $N_\mathrm{extra}=2$.  To
evaluate the sensitivity of the outcome to $N_\mathrm{extra}$, we ran
two $128^2$ disk models with $\mathtt{TOL}=10^{-6}$: one with no extra
iterations and the other with $N_\mathrm{extra}=2$.  The quantities
$\dot{L}/\dot{M}_\circ$, $\dot{E}/\dot{M}_\circ$,  and $\dot{M}_\circ$
deviate between the two runs by less $4\%$ at any given time, far less
than what is seen between two otherwise identical runs at resolutions of
$128^2$ and $256^2$.  We conclude that the inversion error is in some
mean sense much less than the discretization error when \texttt{TOL} is below
$10^{-6}$.   

The $E_\mathrm{NR}$ distribution of the $N_\mathrm{extra}=0$ run is,
however, significantly different than the one shown in
Figure~\ref{fig:errx-disk}.   Without the extra iterations, the
$E_\mathrm{NR}$ distribution is approximately uniform over $10^{-15} <
E_\mathrm{NR} < 10^{-6}$, and becomes more similar to the
$N_\mathrm{NR}=2$ distribution for $E_\mathrm{NR}<10^{-6}$.  The extra
iterations therefore seem to be doing what we expected for the  most
part.  In addition, using two extra iterations per inversion only
increases the simulation's run-time by $4\%$.  Even though this doubles
the run-time contribution from the primitive variable solver, it is
still an insignificant portion of the evolution's total computational
cost, which seems worthwhile to nearly eliminate the possibility that
inversion error makes a significant contribution to the computational
error budget.

This investigation highlights the importance of handling aberrant cells
in real applications.  When an inversion fails to converge to a
solution, when a solution leads to unphysical primitive variables, or
when it converges to a set of primitive variables which we have some {\it a
priori} reason for classifying as numerical artifacts, one must either
halt the run or else ``correct'' these values\footnote{In accretion
disk simulations we typically classify a point as aberrant if 
$\rho_\circ,u \le 0$,  $\gamma > 50$, or $\gamma < 1 $.}.  For
example, the most common problem involves evolving to a state of
negative internal energy density.  We have found that using an
artificial floor can lead to the generation of spontaneous
``explosions.'' These explosions can corrupt and eventually terminate
the evolution.  We have found that $2^\mathrm{nd}$-order interpolation
of $\mathbf{P}$ from the problematic cell's nearest-neighbors is
successful at eliminating these cell-scale artifacts.  This
interpolation procedure is used whenever at least one of the above
conditions is met.  All disk evolutions mentioned in this paper used
this method. 

\section{Conclusion}
\label{sec:conclusion}

We have outlined a compact derivation of the equations for calculating
$\mathbf{P}(\mathbf{U})$.  This formulation suggests a variety of
possibilities for performing the calculation numerically.  We have
implemented a small subset of the possible methods, and compared these
implementations through (1) a survey over a large subset of possible
$\bU$ values; (2) embedding the inverter in a SRMHD code and evolving a
cylindrical explosion problem due to Komissarov; (3) embedding the
inverter in a GRMHD code and evolving a turbulent, magnetized disk
around a rotating black hole.

Several key points emerge from the comparison.  First, the
implementation can be made accurate enough that variable inversion does
not make a significant contribution to the code error budget.  Second,
some implementations can be made fast enough that they occupy only a few
percent of the total cycles used by our GRMHD code.  Since we are using
a particularly simple GRMHD algorithm, this is likely an upper limit to
the fractional cost of the inversion in all GRMHD schemes.  Third, 
the inversion routine originally used in HARM \citep{harm}, called
``5D'' here, is particularly slow and inaccurate.

We recommend the ``2D'' scheme, described in \S 3.2.  A version of this
scheme is available for download on the web at {\tt
http://rainman.astro.uiuc.edu/codelib/codes/pvs.tgz}

\acknowledgments

This work was supported by NSF grants PHY 02-05155, AST 00-93091 and PHY99-07949 (Kavli
Institute for Theoretical Physics, where a portion of this work was completed).
We thank Yuk Tung Liu, Po Kin Leung and Xiaoyue Guan for comments.

\clearpage

\begin{deluxetable}{lccc}
\tablewidth{0pt}
\tablecaption{Physical Solutions for W of the \OneDVsqOrig Scheme's Cubic 
\label{table:phys-solutions}}
\tablehead{\colhead{Case}  &\colhead{Condition} &\colhead{$\mathcal{D}$} 
&\colhead{$W$\tablenotemark{a}} }
\startdata
1 & $d_2^3 > d_0 > 0$ & $\mathcal{D} < 0$  & (\ref{simp-cubic-sol}) \\[0.25cm]
2 & $d_2^3 < d_0 > 0 \ , \ d_2 \ne 0$ & $\mathcal{D} > 0$ & (\ref{gen-cubic-sol}) \\[0.25cm]
3 & $d_2 = 0 \ , \ d_0 > 0 $ & $\mathcal{D} > 0$  &$W = \left( 4 d_0 \right)^{1/3}$ \\[0.25cm]
4 & $d_2^3 = d_0 > 0$ & $\mathcal{D} = 0$  & $W = d_2$ \\[0.25cm]
5 & $d_0 = 0 \ , \ d_2 \ne 0 $ & $\mathcal{D} = 0$ &$W = - 3 d_2$  iff $d_2<0$ \\[0.25cm]
6 & $d_2 = d_0 = 0 $ & $\mathcal{D} = 0$  & (none)\tablenotemark{b}  \\
\enddata
\tablenotetext{a}{Only physically-acceptable solutions, i.e. those that are real and positive, are given here.}
\tablenotetext{b}{The only solution, $W=0$, is unphysical.}
\end{deluxetable}

\begin{deluxetable}{ccc}
\tablewidth{0pt}
\tablecaption{Kerr-Schild Coordinates ($a=0.9375$) used in the Parameter Space Survey \label{table:pspace-points}}
\tablehead{\colhead{$\cos(\Phi)$\tablenotemark{a}}  
&\colhead{$r$}  &\colhead{$\theta$}  }  
\startdata
-0.751 &     8.195 &     1.552  \\ 
-0.250 &     1.375 &     1.444  \\ 
-0.500 &     2.676 &     1.016  \\ 
 1.000 &    23.166 &     2.672  \\ 
-0.997 &    26.467 &     0.658  \\ 
 0.500 &     1.571 &     1.589  \\ 
 0.749 &     3.588 &     1.455  \\ 
 0.250 &     2.406 &     2.483  \\ 
-0.0005 &    35.480 &     0.146  \\ 
\enddata
\tablecomments{Please refer to the electronic version of the Journal for a complete, 
machine-readable table of the parameters used for the survey.  The larger table
provides $x^{(1)},  x^{(2)},  r,  \theta,   \tilde{u}^i, B^j, g_{\mu \nu}$ 
for each value of  $\cos \Phi$.  Please refer to Section~\ref{sec:accr-disk-evol} for 
a definition of the coordinates. }
\tablenotetext{a}{$\cos(\Phi) = \frac{\tu_i B^i}{\sqrt{\tu_i \tu^i B_j B^j}}$}
\end{deluxetable}

\begin{deluxetable}{cccc}
\tablewidth{0pt}
\tablecaption{Parameter Space Efficiency Comparison \label{table:pspace-efficiency}}
\tablehead{\colhead{Method}  &\colhead{NR steps per sol.}  &\colhead{Sol. per sec.} &\colhead{Failure Rate}}  
\startdata
2D            &$8.45$     &$1.66 \times 10^5 $  &$8.7\times 10^{-7}$     \\   
\OneDW        &$7.45$     &$1.68 \times 10^5 $  &$8.8\times 10^{-4}$     \\            
\OneDVsq      &$7.08$     &$1.06 \times 10^5 $  &$3.6\times 10^{-4}$   \\
\OneDVsqOrig  &$7.05$     &$1.24 \times 10^5 $  &$2.0\times 10^{-2}$   \\
5D            &$19.3$     &$1.89 \times 10^4$  &$4.2\times 10^{-1}$   \\ 
Poly          &---      &$9.21 \times 10^3$  &$4.1\times 10^{-2}$   \\
\enddata
\tablecomments{Average number of NR steps taken per solution, average solution rate, and average 
failure rate (per solution) are shown for each method.  The first entry for the polynomial  method is 
vacant since it does not use the NR scheme.  The Intel C Compiler Version 8 and an Intel Xeon 3.06GHz 
workstation were used for these runs.}
\end{deluxetable}

\begin{deluxetable}{cccc}
\tablewidth{0pt}
\tablecaption{Cylindrical Explosion Efficiency Comparison \label{table:cyl-exp-efficiency}}
\tablehead{\colhead{Method}  &\colhead{NR steps per sol.}
  &\colhead{Zone-cycles/sec.} &\colhead{Failure Rate}}  
\startdata
2D            &$3.81$     &$74142$  &$3.75\times 10^{-7}$     \\   
\OneDW        &$3.76$     &$68966$  &$3.75\times 10^{-7}$     \\            
\OneDVsq      &$4.68$     &$58042$  &$3.75\times 10^{-7}$   \\
5D            &$4.45$     &$27100$  &$3.75\times 10^{-7}$   \\ 
\enddata
\tablecomments{Average number of NR steps taken per solution, rate of full zone (cell) updates, and 
average failure rate (per solution) are shown for each method.  The Intel C Compiler Version 8 and 
an Intel Xeon 3.06GHz workstation were used for these runs.}
\end{deluxetable}

\begin{deluxetable}{cccc}
\tablewidth{0pt}
\tablecaption{Comparison of Accretion Rates \label{table:acc-accretion-rates}}
\tablehead{\colhead{Method}  &\colhead{$<\dot{M}_\circ>$} &\colhead{$<\dot{E}/\dot{M}_\circ>$} 
  &\colhead{$<\dot{L}/\dot{M}_\circ>$}}  
\startdata
2D            &$-1.249 $      &$0.86 $      &$1.41 $    \\
\OneDW        &$-1.145 $      &$0.86 $      &$1.32 $    \\
\OneDVsq      &$-1.235 $      &$0.86 $      &$1.43 $    \\
5D            &$-1.226 $      &$0.86 $      &$1.40 $    \\
Thin disk     &---            &$0.82 $      &$1.95 $   
\enddata
\end{deluxetable}

\begin{deluxetable}{cccc}
\tablewidth{0pt}
\tablecaption{Accretion Disk Efficiency Comparison \label{table:acc-efficiency}}
\tablehead{\colhead{Method}  &\colhead{NR steps per sol.}
  &\colhead{Zone-cycles/node/sec. \tablenotemark{a}} 
&\colhead{Failure Rate}}  
\startdata
2D            &$4.19$     &$24535$  &$9.57\times 10^{-5}$     \\   
\OneDW        &$4.18$     &$23860$  &$9.33\times 10^{-5}$     \\            
\OneDVsq      &$5.22$     &$20585$  &$9.46\times 10^{-5}$   \\
5D            &$4.52$     &$14741$  &$9.22\times 10^{-5}$   \\ 
\enddata
\tablenotetext{a}{Four nodes in parallel were used per run.  The rates assume that run-time scales 
linearly with the number of nodes.}
\tablecomments{Average number of NR steps taken per solution, rate of full zone (cell) per processor 
updates, and average failure rate (per solution) are shown for each method.
The Intel C Compiler Version 7.1 and four Intel Xeon 2.40GHz processors were used for these runs.}
\end{deluxetable}

\clearpage
\begin{figure}
\includegraphics[scale=.77]{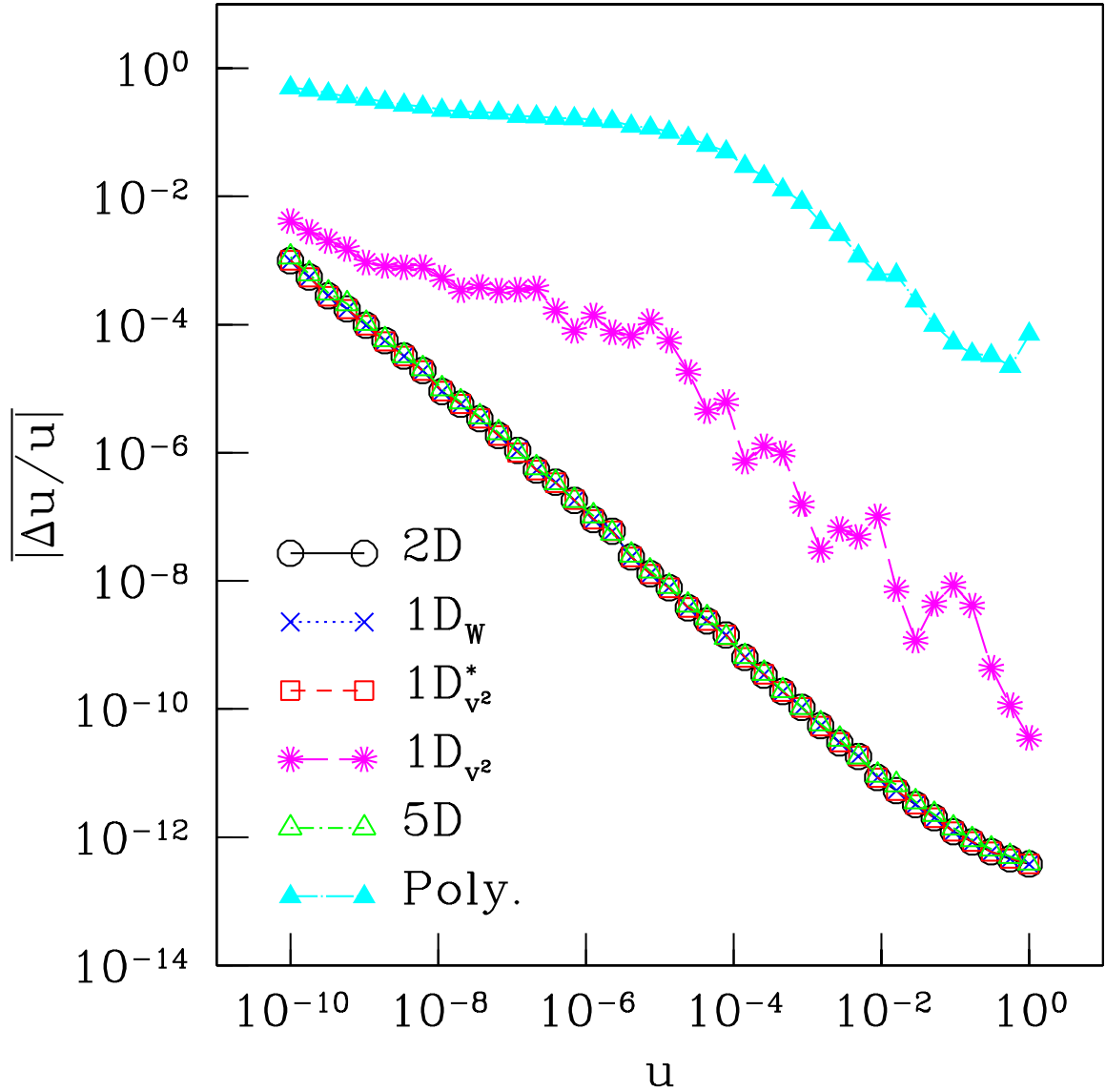}
\caption{
Shown are the relative errors in calculating $u$ from our parameter space survey 
averaged over $\left\{\rho_\circ,B^2,\gamma,\Phi\right\}$ and plotted
versus $\log_{10}(u)$.  Only those points for which all methods found a solution were 
included in the average, accounting for approximately $58\%$ of the surveyed points.  
The curves corresponding to the 2D, \OneDWc, \OneDVsqc, \OneDVsqOrigc, 5D, and polynomial methods are 
represented by, respectively, circles, (blue in the electronic edition) exes, (red) squares, (magenta) 
asterisks, (green) empty triangles, and (cyan) filled triangles; please note 
that the 2D, \OneDWc, \OneDVsqc, and 5D relative errors are almost indistinguishable. 
See the electronic edition of the Journal for a color version of this figure.
\label{fig:rel-error-all}}
\end{figure}

\begin{figure}
\includegraphics[scale=.77]{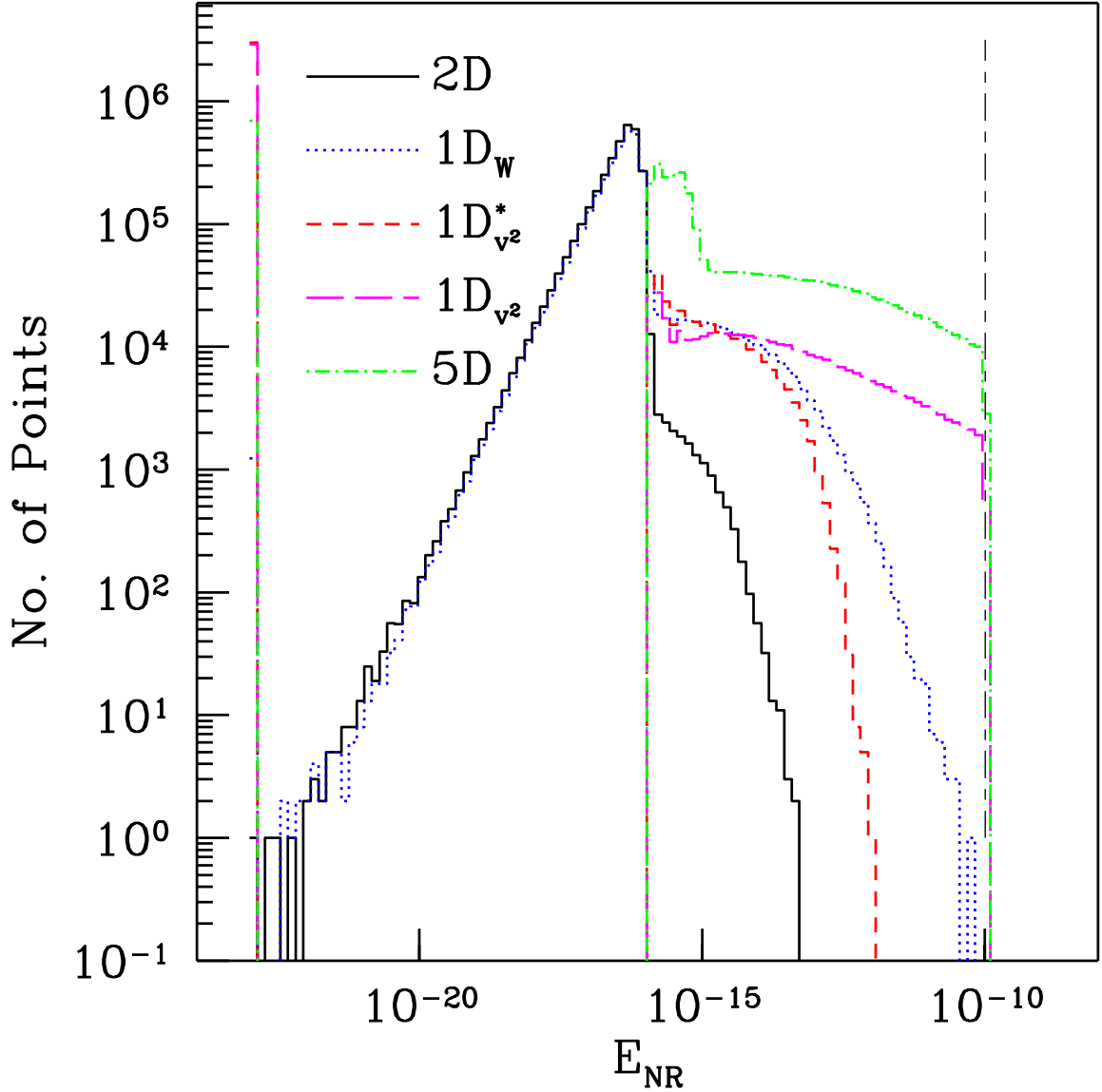}
\caption{Histograms of the final value of $E_\mathrm{NR}$ per parameter space point for the methods 
using a NR scheme.  The distributions generated from the 2D, \OneDWc, \OneDVsqc, \OneDVsqOrig and 
5D methods are represented, respectively, by a solid line, (blue in the electronic edition) dots, 
(red) dashes, (magenta) long dashes, and (green) dot-dashes.  Only those parameter space points for 
which all methods converged to a solution were included in this figure, so all points lie below 
$E_\mathrm{NR} = 10^{-10}$.  There are approximately $3.3 \times 10^6$ points per histogram. 
The first bin contains those points that lie beneath the displayed range.
See the electronic edition of the Journal for a color version of this figure.
\label{fig:errx-pspace}}
\end{figure}

\begin{figure}
\includegraphics[scale=.77]{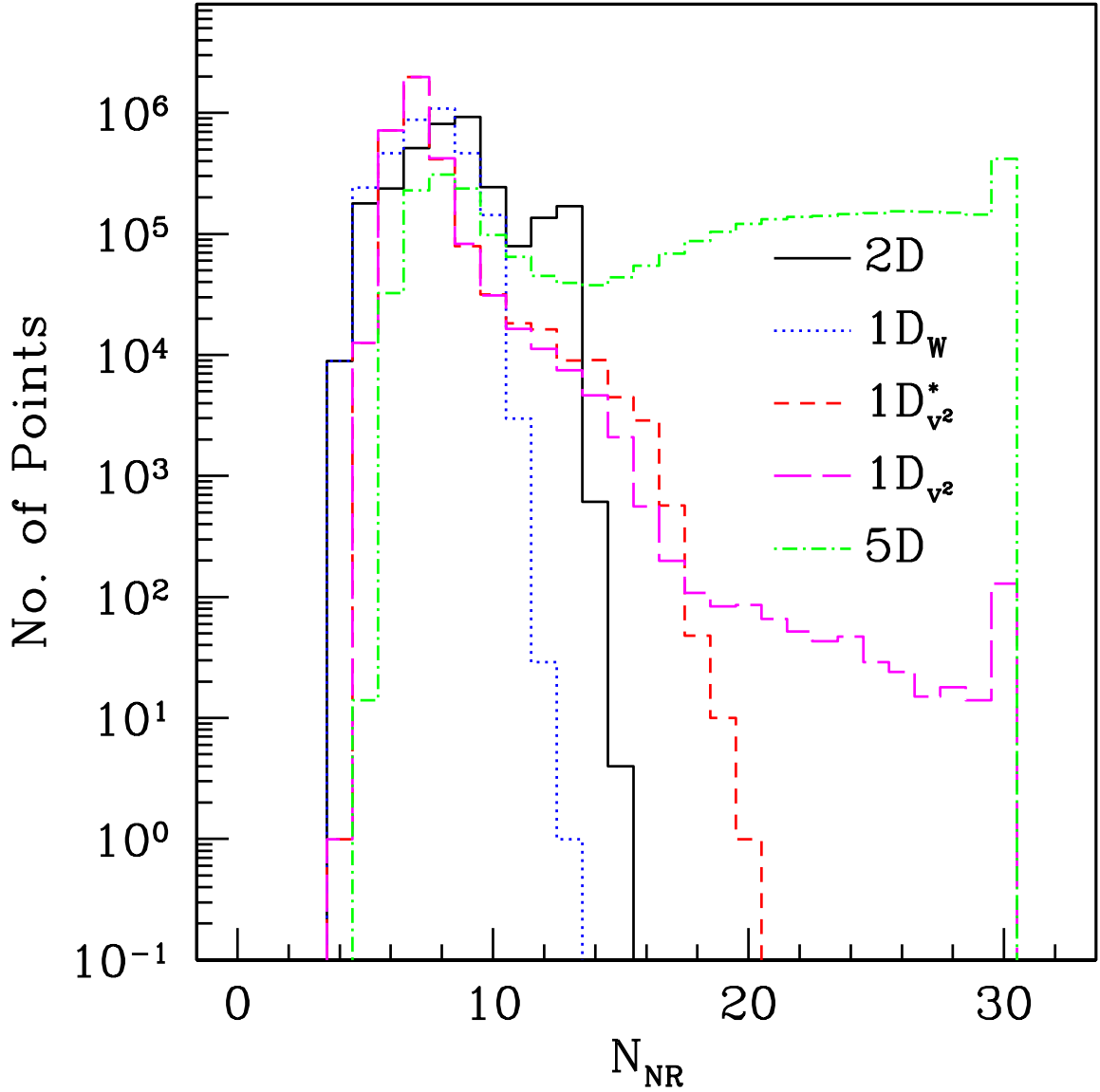}
\caption{Histograms of the number of NR iterations taken by the methods per parameter space point.
The distributions generated from the 2D, \OneDWc, \OneDVsqc, \OneDVsqOrig and 5D methods are 
represented, respectively, by a solid line, (blue in the electronic edition) dots, (red) dashes, 
(magenta) long dashes, and (green) dot-dashes.  Only those parameter space points for which all 
methods converged to a solution were included in this figure.  There are approximately $3.3 \times 10^6$ 
points per histogram.   See the electronic edition of the Journal for a color version of this figure.
\label{fig:niter-pspace}}
\end{figure}

\begin{figure}
\includegraphics[scale=.7]{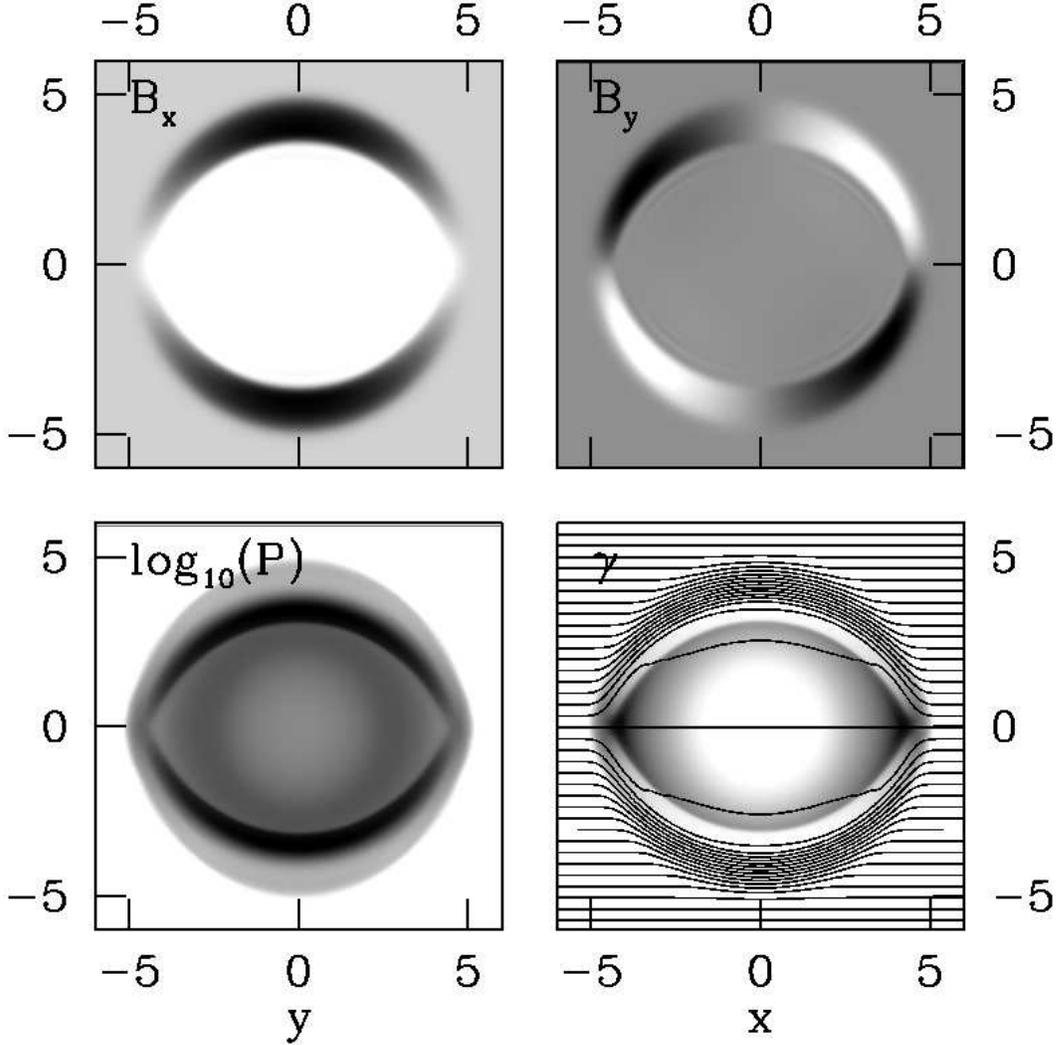}
\caption{
Snapshots of $B_x$ (top left), $B_y$ (top right), $\log_{10} p$ (bottom left) and $\gamma$ 
(bottom right) taken at $t=4$ from the cylindrical explosion evolution.  A continuous greyscale 
from white to black is used for each plot, using the same limits as used by \cite{kom99}:  
$B_x \in [0.008,0.35]$, $B_y \in [-0.18,0.18]$, $\log_{10} p \in [-4.5,-1.5]$, $\gamma \in [1,4.57]$.  
The lower-right plot also shows the magnetic field lines that originate from $x=-6$ and along the $y$-axis 
at equal intervals of $dy=1/3$ from $-5.67$ to $5.67$. 
\label{fig:cyl-exp}}
\end{figure}

\begin{figure}
\includegraphics[scale=.7]{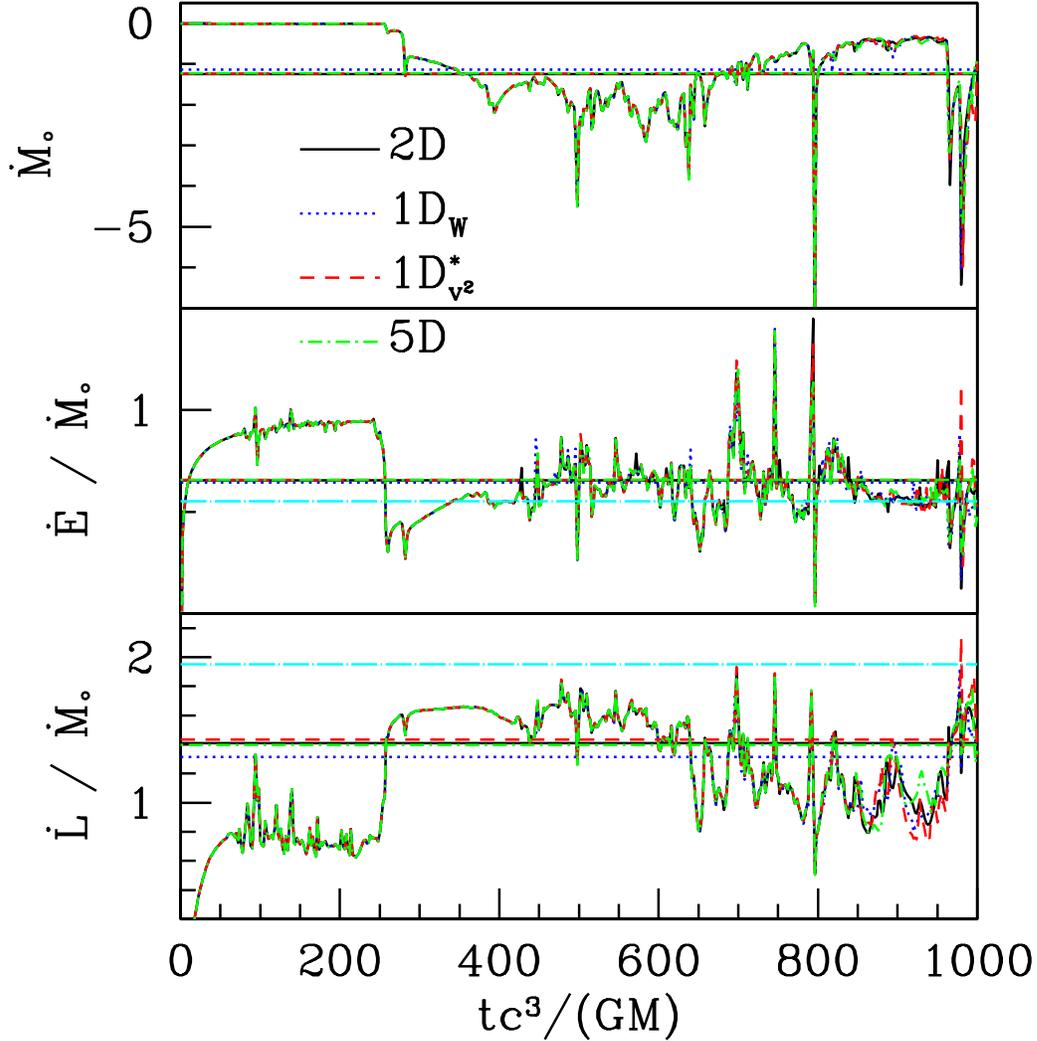}
\caption{
Accretion rates of rest-mass, specific energy and specific angular momentum over time for a 
disk evolution around a Kerr black hole of spin parameter $a=0.9375M$ at a resolution of $256^2$.  
The solid curve, (blue in the electronic version) dots, (red) dashes, and (green) dot-dashes represent 
runs that used---respectively---the 2D, \OneDWc, \OneDVsq and 5D methods. 
All the straight horizontal lines indicate the time-averages of the accretion rates 
over $t=500-2000~GM/c^3$ for the different methods except for the (cyan) lines of dots and long 
dashes which denote the thin disk values.  See the electronic edition of the Journal for a color 
version of this figure.  
\label{fig:accretion-rates}}
\end{figure}

\begin{figure}
\includegraphics[scale=.45]{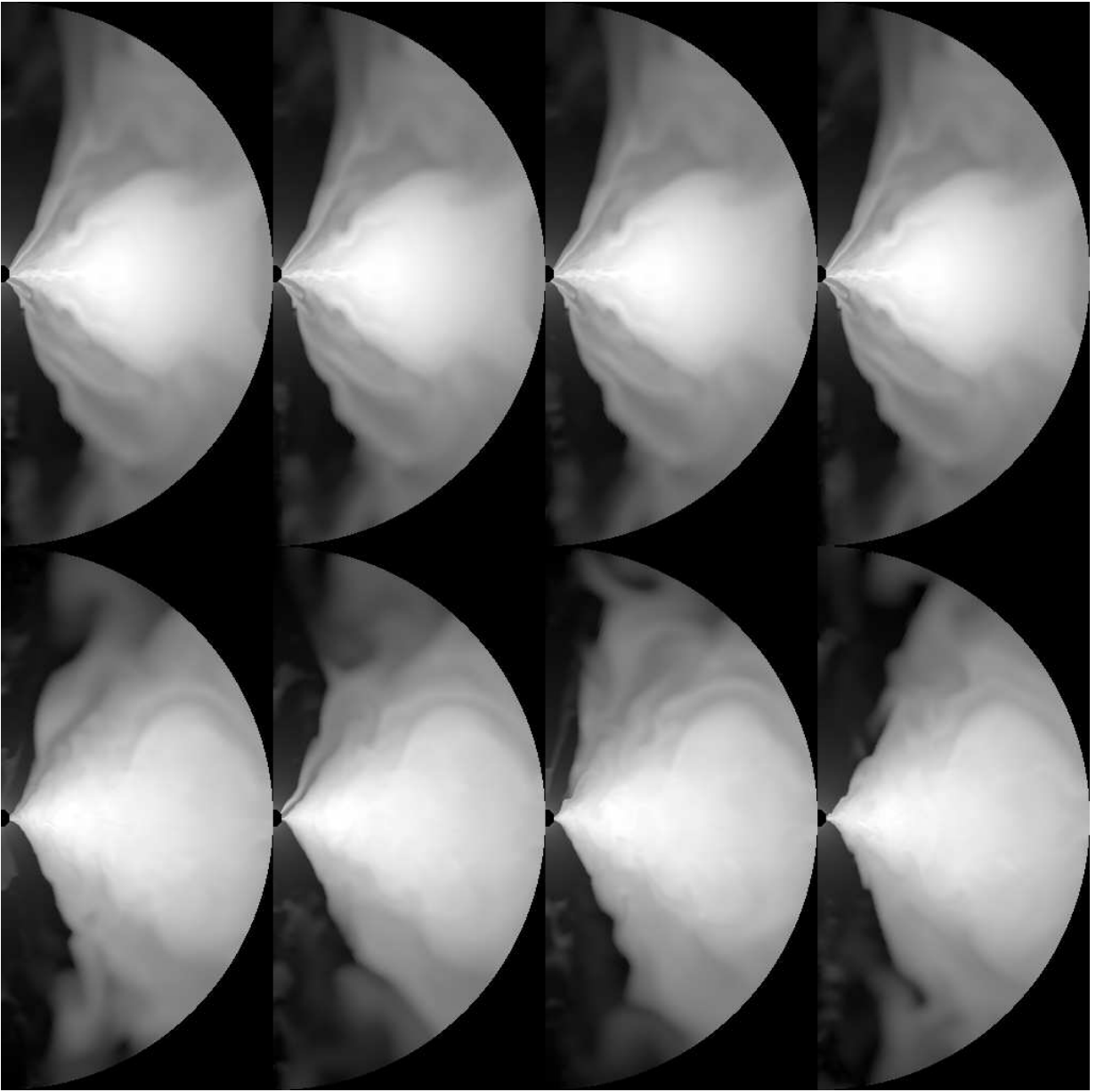}
\caption{Snapshots of $\log \rho_\circ$ at $t=800~GM/c^3$ (top row)
and $t=2000~GM/c^3$ (bottom row) on a $256^2$ grid are shown from accretion disk 
evolutions using different methods:  (from left to right)  the 2D, \OneDWc, \OneDVsq and 5D methods.
The data are plotted here in $(r,\theta)$ coordinates in the $xy$-plane. The region we exclude about 
the origin to excise the singularity from the grid can be seen near the origin.   
The color map is logarithmically-spaced such that the black 
(dark blue in the electronic version) points refer to $\rho_\circ \simeq 4\times10^{-7}$, 
while the white (dark red in the electronic version) regions correspond to $\rho_\circ \simeq 0.69$.
See the electronic edition of the Journal for a color version of this figure.
\label{fig:log-rho}}
\end{figure}

\begin{figure}
\includegraphics[scale=.77]{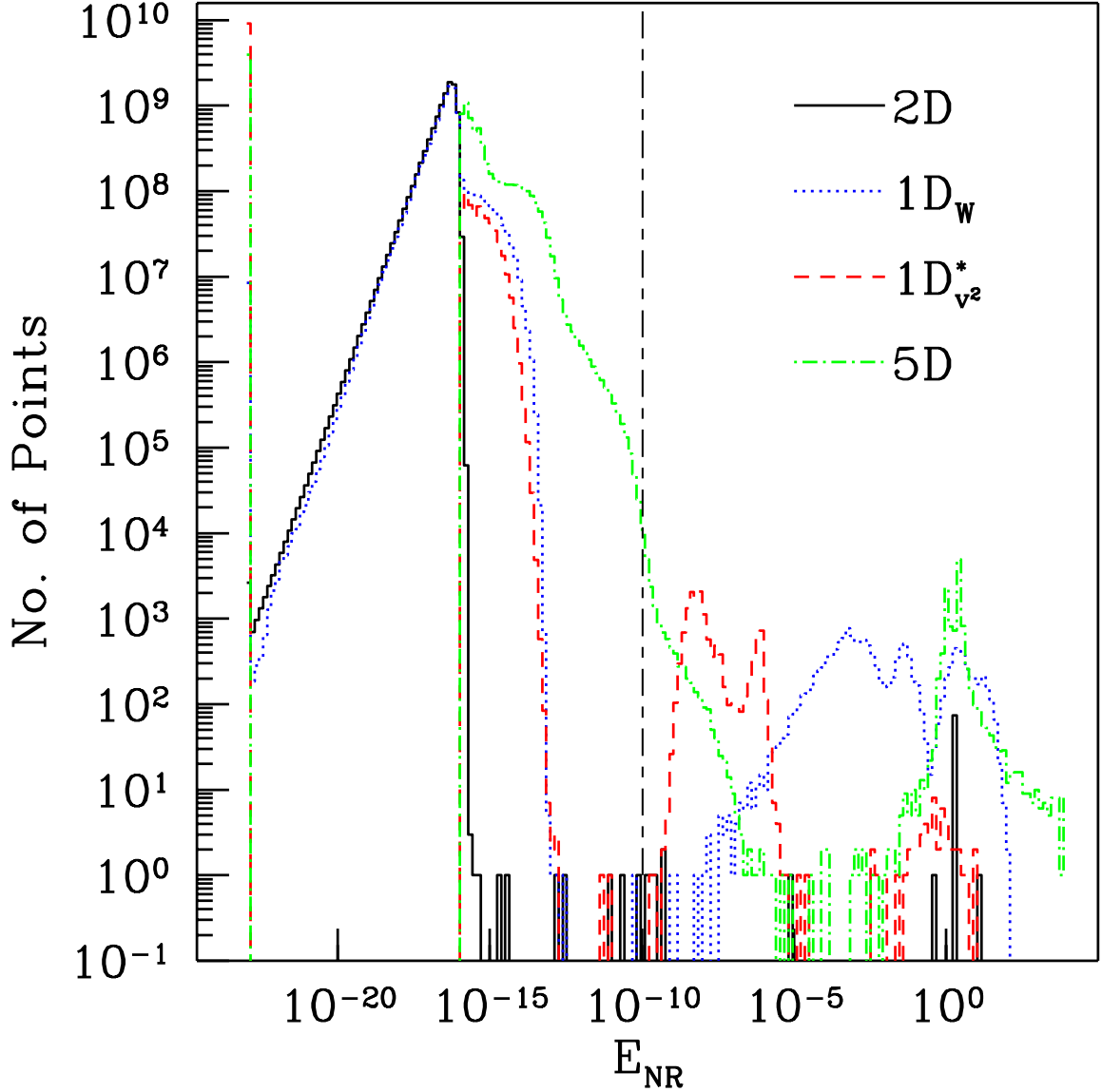}
\caption{Histograms of the values of $E_\mathrm{NR}$ with which the methods return, 
over the entire spacetime region covered by the accretion disk evolution at a resolution of $128^2$.  
The solid curve, (blue in the electronic version) dots, (red) dashes, and (green) dot-dashes 
represent runs that used---respectively---the 2D, \OneDWc, \OneDVsq and 5D methods. Again, 
points with $E_\mathrm{NR}>10^{-10}$ are considered erroneous and are 
replaced by interpolated values during the simulation (see text for further details).
Approximately  $3.8 \times 10^9$ points are represented here.  
The leftmost bin contains those points that lie below the displayed range.
\label{fig:errx-disk}}
\end{figure}

\begin{figure}
\includegraphics[scale=.77]{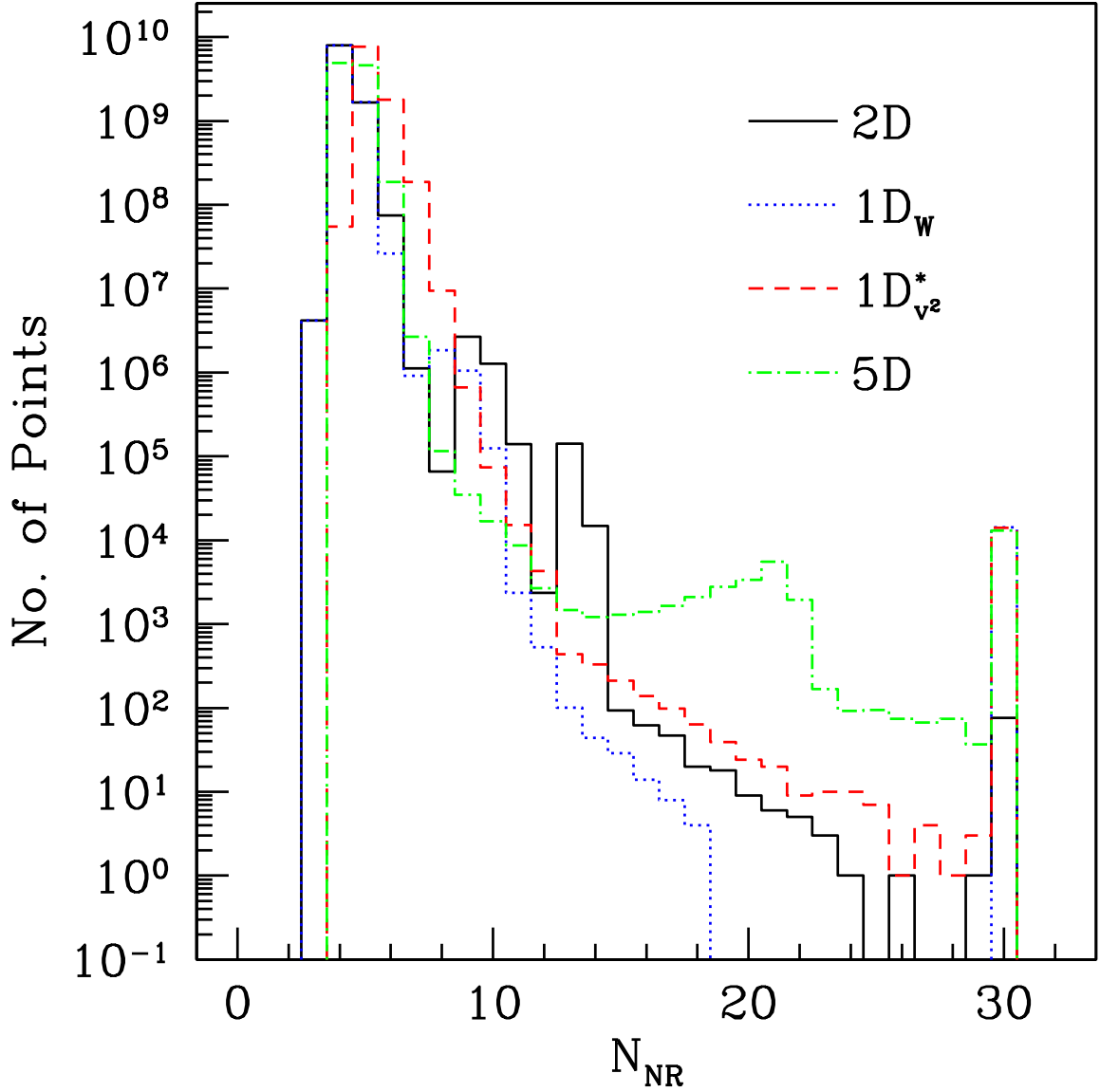}
\caption{Histograms of the number of NR iterations taken by the methods over the entire 
spacetime region covered in the accretion disk evolution at a resolution of $128^2$.
Approximately  $3.8 \times 10^9$ points are represented here.  The bin located 
at $N_\mathrm{NR}=31$ contains all those 
points for which no solution was found, all other points were successful.
\label{fig:niter-disk}}
\end{figure}

\end{document}